\documentclass[11pt]{article}
\usepackage{putex}
\usepackage[pdftex]{graphicx}

\begin{document}
\preprint{COLO-HEP-563 \\ PUPT-2388}

\institution{CU}{${}^1$Department of Physics, 390 UCB, University of Colorado, Boulder, CO 80309, USA}
\institution{PU}{${}^2$Joseph Henry Laboratories, Princeton University, Princeton, NJ 08544, USA}

\title{Dynamic critical phenomena \\ at a holographic critical point}

\authors{Oliver DeWolfe,${}^\CU$ Steven S.~Gubser,${}^\PU$ and Christopher Rosen${}^\CU$}

\abstract{We study time-dependent perturbations to a family of five-dimensional black hole spacetimes constructed as a holographic model of the QCD phase diagram.  We use the results to calculate two transport coefficients, the bulk viscosity and conductivity, as well as the associated baryon diffusion constant, throughout the phase diagram.  Near the critical point in the $T$-$\mu$ plane, the transport coefficients remain finite, although their derivatives diverge, and the diffusion goes to zero.  This provides further evidence that large-$N_c$ gauge theories suppress convective transport.  We also find a divergence in the low-temperature bulk viscosity, outside the region expected to match QCD, and compare the results to the transport behavior of known R-charged black holes.}

\date{August 2011}

\maketitle

\tableofcontents

\section{Introduction and Summary}

The phase diagram of quantum chromodynamics (QCD) is of great interest.  At vanishing chemical potential for baryon number, the theory is expected to evolve from a gas of hadrons at low temperature to a plasma of liberated quarks and gluons at high temperature.  Were chiral symmetry exact, this evolution would progress through a true phase transition; given the breaking of chiral symmetry by quark masses, however, it is predicted by lattice gauge theory instead to be a smooth but rapid crossover (for recent lattice results, see  \cite{Karsch:2007dp,Bazavov:2009zn,Borsanyi:2010cj}).  The results of heavy ion experiments at the Relativistic Heavy Ion Collider (RHIC) and now at the Large Hadron Collider (LHC) are consistent with such a picture (see for example \cite{Jacobs:2004qv}).  However, near the critical temperature, a gas of quarks and gluons is not observed: instead, the physics is well-described by the hydrodynamics of a near-perfect fluid, with no apparent quasiparticle description.  Such a fluid is a good candidate to be modeled by a gravity dual via the AdS/CFT correspondence (for reviews, see \cite{Liu:2007ab,Gubser:2009md,CasalderreySolana:2011us}), and indeed the observed very small ratio of shear viscosity to entropy density \cite{Policastro:2001yc} is a generic feature of such models \cite{Buchel:2003tz,Kovtun:2004de}.

\begin{figure}
\begin{center}
\includegraphics[scale=0.3]{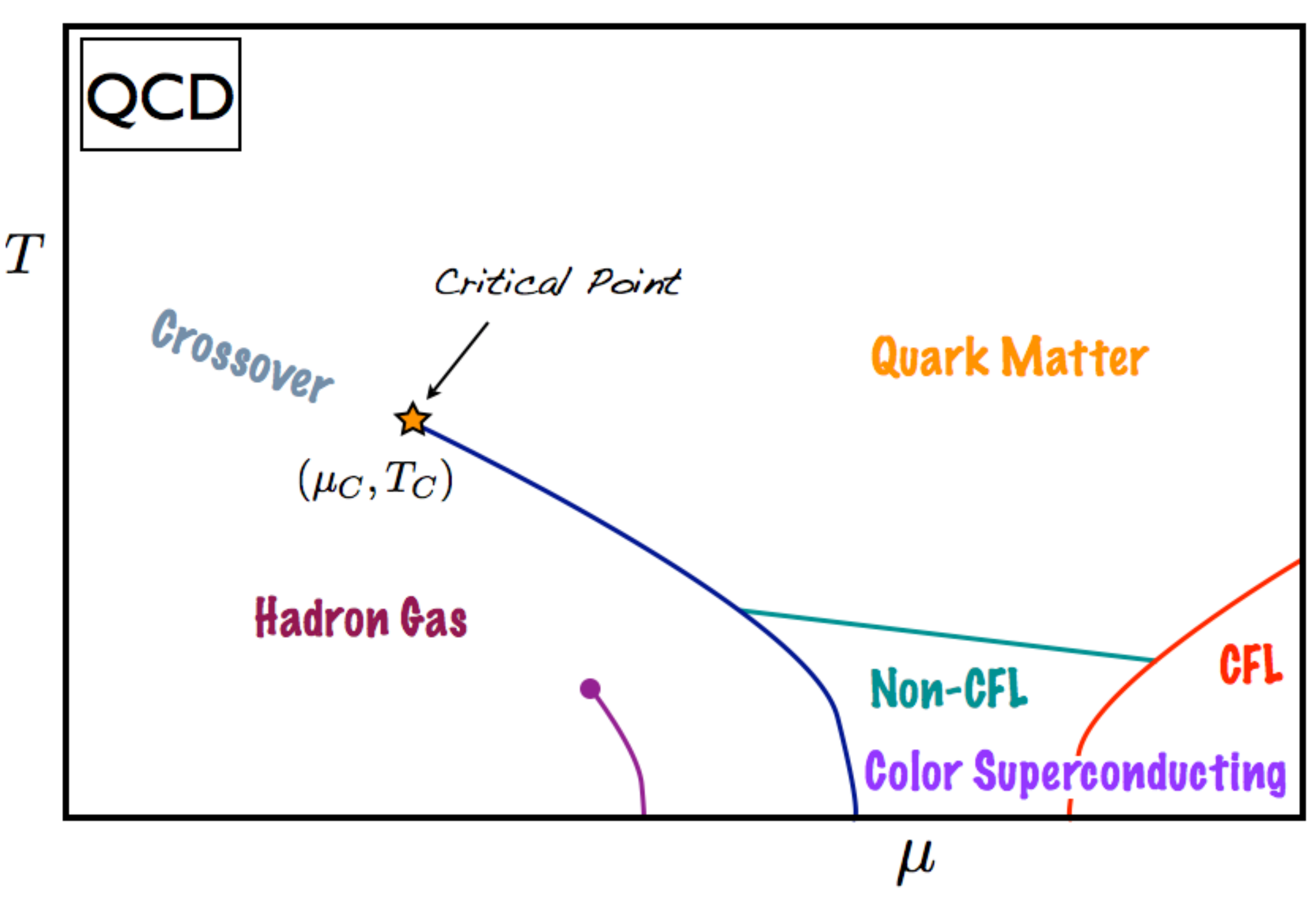}
\caption{The expected phase diagram of QCD.  The line ending in a star is the first-order chiral transition and its critical endpoint. Below is the nuclear matter transition.  At lower right are color superconducting phases, color-flavor locked and otherwise.
\label{PhaseDiag}}
\end{center}
\end{figure}

For QCD at nonzero chemical potential, it is expected that the crossover sharpens into a  line of true first-order phase transitions, ending on a critical point.  The temperature-baryon chemical potential ($T$-$\mu$) phase diagram is illustrated in figure~\ref{PhaseDiag}; for reviews see \cite{Kogut:2004su,Stephanov:2007fk, Alford:2007xm}.  Based on dimensionality and symmetry, this critical point is expected to be in the universality class of the three-dimensional Ising model, like the liquid-gas transition of an ordinary fluid.  Although it has not been accessed experimentally so far, the critical point could be probed in future heavy ion experiments at RHIC, LHC or the proposed fixed-target Compressed Baryonic Matter (CBM) project at the Facility for Antiproton and Ion Research (FAIR) \cite{Aggarwal:2010cw,Staszel:2010zz}.  It is difficult, however, for lattice gauge theory techniques to be used at nonvanishing chemical potential, due to the sign problem of the fermion determinant.  While various lattice techniques such as reweighting \cite{Fodor:2001pe,Allton:2002zi} and imaginary chemical potential \cite{deForcrand:2002ci} have been employed, and other methods such as Nambu--Jona-Lasinio models have been used \cite{Asakawa:1989bq, Scavenius:2000qd, Berges:1998rc},  another theoretical approach would be welcome, and the AdS/CFT correspondence offers such an approach.

In \cite{DeWolfe:2010he}, we constructed a holographic model of the QCD critical point.  A bottom-up method was used: the gravity dual was constructed using the minimal set of fields that captured the essential dynamics.  These fields were a metric tensor characterizing the geometry, a scalar field to encode the running of the gauge coupling, and a gauge field dual to the $U(1)$ current for baryon number, to provide the chemical potential; note that from the AdS/CFT point of view, going to nonzero chemical potential involves merely adding one new field.  Points in the phase diagram correspond to black hole solutions with varying Hawking temperature and electric charge.

The resulting five-dimensional gravitational theory contained some freedom: the scalar potential and the gauge kinetic function, both of which were arbitrary functions of the scalar field.  To capture the behavior of QCD, these functions were chosen to make the zero-chemical potential thermodynamics, including quark number susceptibility, match the crossover behavior predicted by lattice QCD, building on work in \cite{Gubser:2008ny,Gubser:2008sz}; a related body of work includes \cite{Gursoy:2008za,Gursoy:2009kk} and the review \cite{Gursoy:2010fj}.  While the potentials used in our work do not come from any known string theory construction, they were chosen as combinations of exponentials to match the qualitative features of five-dimensional supergravity.

Having thus completely constrained the theory by matching to lattice QCD on the $T$-axis, ensembles of charged black holes were obtained numerically to fill out the phase diagram in the $T$-$\mu$ plane.  As anticipated, the crossover indeed sharpens into a line of first-order phase transitions at finite $\mu$; black holes corresponding to both phases as well as the intermediate thermodynamically unstable configuration were found.  Moreover, the static critical phenomena of the critical point were studied, and the critical exponents were found to be consistent with the Ising mean field values $(\alpha, \beta, \gamma, \delta) = (0, 1/2, 1, 3)$.  The location of the critical point in this model was predicted to be $(T_c, \mu_c) =$ (143 MeV, 783 MeV).  Thus a realistic, consistent holographic realization of the low-$\mu$ QCD phase diagram emerged.\footnote{Expected color superconducting phases at high chemical potential cannot appear in this simple model, and are an interesting task for the future; for progress see for example \cite{Basu:2011yg}.}

It is interesting to move beyond static phenomena, and to study dynamics.  Time-dependent critical behavior includes transport properties, relaxation times and the response to time-dependent perturbations, all necessary for the understanding of a heavy ion collision evolving near the critical point.  There are three relevant transport coefficients: the shear and bulk viscosities associated to the traceless and trace perturbations of the energy-momentum tensor, $\bar\eta$ and $\zeta$ respectively,\footnote{In keeping with the literature, we denote the shear viscosity $\bar\eta$ to avoid confusion with the critical exponent $\eta$.} and the conductivity $\lambda$ of the $U(1)$ baryon density.  The ratio of shear viscosity to entropy density is known to be universal  $\bar\eta/s = 1/4 \pi$ in all two-derivative gravity models \cite{Buchel:2003tz,Kovtun:2004de}, but $\lambda$ and $\zeta$ are expected to depend nontrivially on the location in the phase diagram.  Moreover, the dynamic critical exponent $z$ controls
the phenomenon of critical slowing down, where 
the equilibration time $\tau$ grows with the correlation length $\xi$   as $\tau \sim \xi^z$; this behavior is expected to determine how close to criticality a heavy ion collision can approach in the limited time available before freeze-out and hadronization, as $\xi < \hbox{(time)}^{1/z}$ \cite{Stephanov:1999zu,Berdnikov:1999ph}.

Static critical phenomena are sorted into universality classes by properties such as their dimensionality and symmetry of the order parameter.  This concept was extended to dynamic critical phenomena by Hohenberg and Halperin \cite{Hohenberg:1977ym}, who classified various universal ``models" based on their conserved quantities, which manifest as hydrodynamic modes, and the Poisson brackets between them and with the order parameter; dynamic universality classes are then determined by these models as well as the static class.  Son and Stephanov argued that in QCD, only one combination of the baryon density and the chiral condensate survives as a hydrodynamic mode, and thus having this single conserved mode as well as conserved energy-momentum, the theory should fit into dynamic model H \cite{Son:2004iv}.

The response of a fluid to charge inhomogeneities is controlled by the diffusion constant $D$, associated to the dispersion relation $\omega  = -i D k^2$.  The diffusion constant is related to the conductivity $\lambda$  and the charge susceptibility $\chi$ by
\eqn{}{
D= {\lambda \over\chi} \,,
}
and at a critical point, $D$ tends to zero while $\chi$ diverges. The behavior of $\lambda$ near the critical point, however, depends on the dynamic universality class.  Hydrodynamic models neglecting nonlinear interactions predict that $\lambda$ remains finite at criticality; this is characteristic of Hohenberg and Halperin's model B, where the single hydrodynamic mode is taken to be a conserved density.  The only kind of conduction possible in such a model is via diffusion.  Model B predicts the dynamic critical exponent
\eqn{ModelBz}{
z = 4 - \eta \,,
}
where $\eta$ is the usual static exponent giving the anomalous dimension of the density two-point function.   In model H, on the other hand, the inclusion of energy and momentum as hydrodynamic modes makes convective conductivity possible, and this new channel naively dominates.  In model H, the conductivity $\lambda$ and shear viscosity $\bar\eta$ diverge at the critical point,
\eqn{}{
\lambda \sim |T - T_c|^{x_\lambda} \,, \quad \quad
\bar\eta \sim |T - T_c |^{x_\eta} \,,
}
with exponents $x_\lambda$ and $x_\eta$ which are related to the static exponent $\eta$ by
\eqn{}{
x_\lambda + x_\eta = 4 - d - \eta \,,
}
where $d$ is the spatial dimensionality.  In the 3D Ising model $\eta$ and $x_\eta$ are close to zero, giving $x_\lambda$ close to one \cite{Son:2004iv}.  The critical exponent $z$ when convection dominates takes the value
\eqn{}{
z = 4 - \eta - x_\lambda \,,
}
and is thus moved from $z \approx 4$ in model B to $z \approx 3$ in model H.  Note that even in models where $\lambda$ diverges, $\chi$ always diverges faster, so $D$ still goes to zero \cite{Hohenberg:1977ym}.  

The expected behavior of bulk viscosity near the liquid-gas critical point was investigated by Onuki \cite{Onuki:1997}, who predicted a divergence also depending on $z$,
\eqn{Onuki}{
\zeta \sim |T - T_c|^{-z \nu + \alpha}
}
where $\nu>0$ and $\alpha$ are static exponents; for several other predictions for bulk viscosity at a critical point, see for example \cite{Buchel:2009mf}.

Work has been done applying the AdS/CFT correspondence to dynamic critical phenomena in various models.  Bulk viscosity for the models we consider has been computed at vanishing chemical potential in \cite{Gubser:2008yx, Gubser:2008sz}.  Maeda, Natsuume and Okamura calculated the conductivity in the ``one-charge black hole" model dual to ${\cal N}=4$ Super-Yang Mills with a chemical potential---hereafter the one-charge ${\cal N}=4$ black hole---and found it to stay finite at the critical point, identifying the result with model B despite the presence of conserved energy and momentum \cite{Maeda:2008hn}.  The natural speculation is that the nonlinear interactions responsible for the convective component in the conductivity in model H are suppressed by the large number of colors $N_c$, and in \cite{Natsuume:2010bs} Natsuume and Okamura argued that large-$N_c$ counting indeed enhances the diffusive over the convective conductivity, reducing model H to an effective model B.  Meanwhile, the bulk viscosity was studied in a mass deformation of the one-charge  ${\cal N}=4$ black hole by Buchel \cite{Buchel:2010gd} and in the ${\cal N}=2^*$ model by Buchel and Pagnutti \cite{Buchel:2007mf,Buchel:2008uu}.  In both cases the bulk viscosity was also found to be finite at the critical point, in contradiction with expectations from (\ref{Onuki}).   It is natural to speculate that again the large-$N_c$ limit suppresses the divergence in the transport coefficient.  The dynamic critical exponent $z$ has also been studied directly, yielding the mean-field model B value $z=4$ for the deformed one-charge black hole \cite{Buchel:2010gd} and $z=0$ for ${\cal N}=2^*$ \cite{Buchel:2010ys}.

Dynamic universality classes depend not just on the conserved quantities and their Poisson brackets, but also on the static universality class.  The  work described in the previous paragraph shares the  list of conserved quantities with QCD, but differs in the static critical exponents.  It is natural to wish to extend the study of holographic dynamic critical phenomena to a model that shares the static universality class with QCD as well.
In this paper, we begin the investigation of the dynamic critical phenomena in the QCD-like holographic critical point of \cite{DeWolfe:2010he}, 
 by studying finite-$\omega$ fluctuations around those black hole backgrounds.  With these fluctuations, we are able to calculate the transport coefficients $\zeta$ and $\lambda$, and the associated diffusion $D$.

In keeping with the other AdS/CFT cases, we find the transport coefficients remain finite at the critical point, and the diffusion constant goes to zero.  Thus these models share the apparent suppression of the convective contribution to transport by the large-$N_c$ limit \cite{Natsuume:2010bs}, and behave effectively as model B.  

Furthermore, the behavior of $\lambda$ and $\zeta$ as functions of $T$ and $\mu$ near the critical point is quite similar to that of the 
entropy and baryon densities $s$ and $\rho$: all are smooth approaching the critical point in the $T$-$\mu$ plane along the axis defined by the first-order line, but develop infinite slope when approaching off-axis.  Similar behavior arises in the one-charge ${\cal N}=4$ black hole, where the critical exponent controlling all these divergent slopes is the same,
\eqn{}{
\lambda - \lambda_c \sim s - s_c \sim \rho - \rho_c \sim |T - T_c|^{1/\delta} \,,
}
with $\delta = 2$; the corresponding exponent in the QCD-like case, $\delta \approx 3$, is consistent with our results for $\lambda- \lambda_c$ and $\zeta - \zeta_c$.  This suggests that in the vicinity of the critical point, the deviations of the conductivity and bulk viscosity from their critical values can be thought of as depending smoothly on the deviations of the densities from theirs. 

From these results we can estimate the dynamic critical exponent $z$ assuming a mean-field value of $\eta$; a precise determination of $z$ and $\eta$ requires finite-$k$ fluctuations, which we leave for future work.  Given the model B behavior and the mean field exponents, using the mean field value $\eta = 0$ in \eno{ModelBz}, we can estimate
\eqn{}{
z \approx 4 \,.
}
Another feature of the behavior of the transport coefficients in the phase diagram is worth pointing out.  In  \cite{DeWolfe:2010he} and \cite{Gubser:2008ny}, the potential and gauge kinetic function in the Lagrangian were only constrained to lattice QCD results over certain ranges of temperature, corresponding to certain values of the scalar field.  In principle, one could imagine modifying these functions such that the matched region is not affected, while results change elsewhere; thus the application of these models to QCD should only be trusted in a certain band of temperatures.  This is not unreasonable, since the gravity dual picture is only expected to apply in a certain range near the critical temperature where there is no quasiparticle description.

However, phenomena outside the region designed to match QCD may be interesting in their own right.  We find that as the temperature decreases, both the bulk viscosity and conductivity begin to rise.   In fact, the bulk viscosity has a divergence at a temperature around half the crossover temperature on the $T$-axis, outside the region matched to QCD, and this divergence extends out into the plane.  Unlike the phenomena at the critical point, this divergence in the transport coefficient is not associated with any feature in the thermodynamics of $\rho$ or $s$.  From the gravity point of view, it can be understood as the place where the fluctuation solutions develop a node, but the field theory interpretation is unclear.  Whether the conductivity also has a pole at a still lower temperature is outside the range accessible to our numerical solutions.  We note that a similar divergence appears for the conductivity of the thermodynamically unstable branch of the one-charge black hole, as the ``superstar" limit \cite{Myers:2001aq} is approached.

The summary of the remainder of the paper is as follows.  In section~\ref{BlackHoleSec} we discuss the Lagrangian, equations of motion and gauge symmetries of our class of models, and construct the gauge-invariant fluctuations around the black hole backgrounds.  The fluctuation equations and the Kubo formulae for extracting the transport coefficients are presented in section~\ref{TransportSec}.  In section~\ref{OneChargeSec}, we review the one-charge ${\cal N}=4$ black hole solutions and solve for the conductivity both analytically and numerically, checking our method and pointing out several features that will have analogs in the QCD-like black holes; the analytic calculation was previously worked out in \cite{Maeda:2008hn, Son:2006em}.  In section~\ref{QCDBHSec}, we present the calculations of the transport coefficients for the QCD-like black holes.  We conclude in section~\ref{ConclusionSec}.  Results for the conductivity and diffusion of another model, the two-charge ${\cal N}=4$ black hole, are given in the appendix.

\section{Black hole backgrounds}
\label{BlackHoleSec}

In this section we review the five-dimensional gravity ansatz that applies both to the QCD-like holographic critical models of \cite{DeWolfe:2010he} and the one-charge ${\cal N}=4$ black hole, present the resulting equations of motion and symmetries, and construct the gauge-invariant fluctuations around these backgrounds with nonzero frequency.

\subsection{Lagrangian and ansatz}

We consider five-dimensional gravitational theories containing a metric $g_{\mu\nu}$, a vector field $A_\mu$ and a scalar $\phi$, with Lagrangian
\eqn{LwithF}{
  {\cal L} = {1 \over 2\kappa^2} \left[ 
    R - {f(\phi) \over 4} F_{\mu\nu}^2 - {1 \over 2} (\partial\phi)^2 - V(\phi) \right] \,.
}
The relevant solutions to these theories are
asymptotically AdS black three-brane solutions with radial profiles for the electric potential and scalar field, 
 \eqn{Background}{
ds^2 &= e^{2A(r)} (- h(r) dt^2 + d \vec{x}^2) + {e^{2B(r)} \over h(r)} dr^2 \,, \cr
A_\mu dx^\mu &= \Phi(r)\, dt \,, \quad \quad \phi = \phi(r) \,,
}
and the equations of motion resulting from this ansatz are
 \eqn{SecondOrder}{
  A'' - A' B' + {1 \over 6} \phi'^2 &= 0  \cr
  h'' + (4A'-B') h' - e^{-2A} f(\phi) \Phi'^2 &= 0  \cr
  \Phi'' + (2A' - B') \Phi' + {d\log f \over d\phi} \phi' \Phi' &= 0  \cr
  \phi'' + \left( 4A' - B' + {h' \over h} \right) \phi' - {e^{2B} \over h} 
    {\partial V_{\rm eff} \over \partial\phi} &= 0 \,,
 }
where
 \eqn{VeffDef}{
  V_{\rm eff}(\phi,r) \equiv V(\phi) - {1 \over 2} e^{-2A-2B} f(\phi) \Phi'^2 \,,
 }
along with the zero-energy constraint
 \eqn{ZeroEnergy}{
  h (24 A'^2 - \phi'^2) + 6 A' h' + 2 e^{2B} V(\phi) + e^{-2A} f(\phi) \Phi'^2= 0 \,.
 }
Asymptotically AdS solutions may be written in coordinates that  as $r \to \infty$ approach
\eqn{AdS}{
ds^2 &\to e^{2r/L} \left( -dt^2 + d\vec{x}^2 \right) + dr^2 \,, \cr
\Phi(r) &= \Phi_{(0)} +  \Phi_{(2)} e^{-2 r/L} + \ldots\,,  \cr
\phi(r) &= \phi_{(4-\Delta)} e^{(\Delta-4)r/L} + \ldots + \phi_{(\Delta)} e^{-\Delta r/L} + \ldots \,.
}
Black brane backgrounds have a horizon $r = r_H$ defined by the largest solution to the vanishing of the horizon function $h(r_H) \equiv 0$.  The temperature $T$ and entropy density $s$  can be calculated as
\eqn{TemperatureEntropy}{
T = {1 \over 4 \pi} h'(r_H) e^{A(r_H) - B(r_H)} \,, \quad \quad s = {2 \pi \over \kappa^2} e^{3 A(r_H)} \,,
}
while the chemical potential $\mu$ and $U(1)$ density $\rho$ can be read off from the near-boundary expansion of $\Phi(r)$,
\eqn{MuRho}{
\mu = {\Phi_{(0)} \over L} \,, \quad \quad \rho = -{\Phi_{(2)} \over \kappa^2} \,.
}

\subsection{Gauge symmetries}

The gauge symmetries of the theory are $U(1)$ gauge transformations of the vector field given by $\lambda(x)$, and general coordinate transformations given by $\epsilon^\mu(x)$, with the general transformation of the fields
\eqn{GaugeTrans}{
\delta A_\mu &= \partial_\mu \lambda + \epsilon^\nu \partial_\nu A_\mu + (\partial_\mu \epsilon^\nu) A_\nu \,,
\cr \quad \quad \delta g_{\mu\nu} &= \epsilon^\rho \partial_\rho g_{\mu\nu} + (\partial_\mu \epsilon^\rho) g_{\rho \nu} + (\partial_\nu \epsilon^\rho) g_{\mu\rho} \,, \cr
\delta \phi  &= \epsilon^\mu \partial_\mu \phi \,.
}
To stay consistent with the ansatz \eno{Background}, one may still make transformations with a general $\epsilon^r(r)$ as well as $\epsilon^0(t)$, $\epsilon^i(\vec{x})$ and $\lambda(t)$ obeying
\eqn{ResidualGauge}{
\partial_t \epsilon^0(t) = {\rm const} \,, \quad \quad 
\partial_i \epsilon_j(\vec{x}) + \partial_j \epsilon_i(\vec{x}) = {\rm const} \times \delta_{ij} \,, \quad \quad
\partial_t \lambda(t) = {\rm const} \,.
}
One may use $\epsilon^r(r)$ to choose the metric function $B(r)$ arbitrarily.
In the gauge $B(r)=0$, which we use for the QCD-like holographic critical black holes, the remaining coordinate transformations \eno{ResidualGauge} may be used to set any two of the zero point of $A(r)$ and the overall scales of $h(r)$ and $\Phi(r)$:
\eqn{Rescalings}{
(r, x^i, t) = \left(\gamma \tilde{r}, \beta \tilde{x}^i, {\beta \over \gamma} \tilde{t}\right) \quad \to \quad \tilde{h} = {1 \over \gamma^2} h \,, \quad \tilde{A} = A + \log \beta \,, \quad \tilde{\Phi} = {\beta \over \gamma} \Phi\,,
}
while the gauge transformation $\lambda$ can fix the zero point of $\Phi(r)$.  The remaining transformations are constant shifts of all variables (of which shifts of $t$ and $x^i$ are trivial, while shifts of $r$ will change the form of the various functions), and the manifest $SO(3)$ symmetry acting on $x^i$.

\subsection{Fluctuations and gauge invariant quantities}

We are interested in studying small fluctuations around the background \eno{Background} of the form
 \eqn{Flucts}{
ds^2 &= \Big[g_{\mu\nu}^0 + \Re\{ e^{2A(r)} e^{- i \omega t} h_{\mu\nu}(r) \} \Big]dx^\mu dx^\nu\,, \cr
A_\mu dx^\mu &= \Phi(r)\, dt + \Re\{ e^{ - i \omega t} a_\mu(r) \} dx^\mu  \,, \quad \quad \phi = \phi(r)+\Re\{ e^{ - i \omega t} \tilde\phi(r) \}  \,,
}
where $g_{\mu\nu}^0$ is the background metric, and we have assumed plane-wave dependence in the time direction.  This ansatz preserves the $SO(3)$ rotating the spatial directions together.  From here on, we will omit the instruction to take the real part and work directly with complexified perturbations, on the understanding that appropriate superpositions of complexified perturbations will be real.

Certain combinations of the fluctuations will be gauge degrees of freedom.  These correspond to gauge transformations  at linear order in the small parameter associated to the fluctuations; gauge transformations at zeroth order act on the background.
Gauge transformations preserving the ansatz \eno{Flucts} for the time dependence of the fluctuations also take a plane wave form,
\eqn{}{
\epsilon^\mu(t, \vec{x}, r) = e^{- i \omega t} \xi^\mu(r) \,, \quad \quad
\lambda(t, \vec{x}, r) = e^{- i \omega t} \eta(r) \,.
}
We can then calculate the transformation of the fluctuations $h_{\mu\nu}$, $a_\mu$, $\tilde\phi$ under the gauge transformations.  Here we record only the results for modes with no index in the $r$-direction; all independent fluctuating modes come from this set of fields.  For the metric we have,
\eqn{}{
\delta h_{tt} &=  - (h' + 2 A' h)\xi^r + 2 i \omega h \xi^0 \,, \cr
\delta h_{t i} &= - i \omega \delta_{ij} \xi^j \,, \cr
\delta h_{ij} &= 2 A' \xi^r \delta_{ij} \,,
}
for the gauge field,
\eqn{}{
\delta a_t &= - i  \omega \eta +  \xi^r  \Phi' - i \omega \Phi  \xi^0 \,,\cr
\delta a_i &= 0 \,, 
}
and for the scalar,
\eqn{}{
\delta \tilde\phi = \xi^r \phi' \,.
}
We would like to construct gauge-invariant combinations corresponding to the fluctuating modes of the theory.  We expect a total of nine such fluctuations: five modes from the graviton, three from the gauge field and one from the scalar. 
We immediately see eight such  gauge invariant combinations: the three $a_i$, and the five traceless components of $h_{ij}$.  The trace of the spatial part of the graviton transforms as
\eqn{}{
\delta \left(  {1 \over 3} \left( h_{xx} + h_{yy} + h_{zz}  \right) \right) = 2 A' \xi^r \,,
}
and so we can construct the ninth gauge invariant mode involving the scalar fluctuation,
\eqn{}{
{\cal S} \equiv \tilde\phi  - {\phi' \over 2 A'} {1 \over 3} \left( h_{xx} + h_{yy} + h_{zz}\right) \,.
}
Thus the nontrivial profile for the background scalar field $\phi = \phi(r)$ couples the scalar fluctuations to fluctuations of the graviton trace; in the AdS limit, this reduces to just the scalar fluctuation ${\cal S} \to \tilde\phi$ and the graviton trace becomes a pure gauge mode.  For a nonzero scalar profile, this is the AdS/CFT encoding of the running coupling that breaks scale invariance producing a nonzero trace of the energy-momentum tensor.

While ${\cal S}$ is a natural definition for the fluctuation since it has a smooth AdS limit and shares its possible asymptotic behaviors with $\phi$, it will  be useful for us to consider the rescaled mode
\eqn{HandS}{
{\cal H} &\equiv - {2 A' \over  \phi'} {\cal S} \cr &= {1 \over 3} (h_{xx} + h_{yy} + h_{zz}) - {2 A' \over  \phi'} \tilde\phi \,.
}
This normalization is the natural one for thinking of the mode as the graviton trace; for $\phi' \neq 0 $ one may pick a gauge where $\tilde\phi = 0$, where ${\cal H}$ is the trace precisely.

We see the fields organize themselves under $SO(3)$ as a singlet (the scalar ${\cal S}$ or ${\cal H}$), a triplet (the gauge field $a_i$) and a quintuplet (the traceless graviton $h_{ij}$), and since they are in different representations they cannot mix at the linearized level.  Thus we expect each of these modes to satisfy a decoupled fluctuation equation.  Note this would not hold for a nonzero spatial momentum; instead, a smaller $SO(2)$ symmetry group would organize the perturbations.

\section{Fluctuation equations and transport coefficients}
\label{TransportSec}

The fluctuation equations of the nine modes ${\cal H}$, $a_i$ and $h_{ij}$ can be obtained by linearizing the Einstein, Maxwell and Klein-Gordon equations around the background \eno{Background}.  In general, the various equations that result will also contain the auxiliary quantities $a_r$, $h_{\mu r}$.  By taking linear combinations  of the Einstein and Maxwell equations, these can be eliminated, giving us nine differential equations in nine unknowns.

The coefficients of the fluctuation equations include the background fields $A$, $B$, $h$, $\Phi$ and $\phi$ and their derivatives.  These fields obey the background equations of motion, which act as constraints on these coefficients.  To exhaust the constraints, we use the various background equations \eno{SecondOrder} to eliminate the second derivatives $A''$, $\Phi''$, $h''$ and $\phi''$; in addition, the zero-energy constraint \eno{ZeroEnergy} can be used to eliminate the potential $V(\phi)$, although derivatives of the potential will still appear.  In this way we achieve a unique presentation for each coefficient with no hidden constraints.

Each fluctuation equation can be used to calculate a transport coefficient via a Kubo formula.  These are associated to the imaginary part of the corresponding Green's function, which comes from the ``conserved flux" ${\cal F}$ for each equation.  For a differential equation of the form
\begin{eqnarray}
y''(r) + p(r) y'(r) + q(r) y(r) = 0\,,
\end{eqnarray}
Abel's identity guarantees that for two solutions $y_1(r)$, $y_2(r$), the quantity
\begin{eqnarray}
{\cal F}  \equiv \exp \Big( \int p(r) dr \Big) \, W(y_1, y_2) \,,
\end{eqnarray}
is independent of $r$, where $W(y_1, y_2) \equiv y_1 y_2' - y_2 y_1'$ is the Wronskian of the two solutions.  As long as $p(r)$, $q(r)$ are real, for some complex solution $y(r)$ its conjugate $y^*(r)$ is also a solution.  Taking $y_1 = y$ and $y_2 = y^*$, our expression for the conserved flux becomes
\eqn{AbelFlux}{
{\cal F}  \equiv \exp \Big( \int p(r) dr \Big) \, {\rm Im} \,(y^* y') \,,
}
up to an overall factor.

\subsection{Near-horizon behavior and boundary conditions}

Noting that $h$ has a zero at the horizon, $ h= h'(r_H)  (r - r_H) + \ldots$, while $A$ and its derivatives are regular there, all the  fluctuation equations considered in this paper have the same limit near the horizon,
\eqn{}{
X'' + {1 \over r - r_H} X' + {\omega^2 e^{2B(r_H) - 2A(r_H)} \over h'(r_H)^2 (r - r_H)^2} X = 0 \,,
}
where $X$ is any of $Z$, $a$, ${\cal S}$ or ${\cal H}$ (see the following subsections), and we have assumed that the gauge degree of freedom $B$ is chosen so it and its derivatives do not diverge at the horizon.   Near the horizon, one may expand
\eqn{XSoln}{
X(r) = (r - r_H)^\alpha (x_0 + x_1 (r - r_H) + \ldots) \,,
}
where we must choose the exponent $\alpha$ to be
\eqn{AlphaExponent}{
\alpha = \pm i \omega  {e^{B(r_H) - A(r_H)} \over h'(r_H) } \,.
}
In solving the fluctuation equations we must impose two boundary conditions.  The first is the requirement of infalling boundary conditions at the horizon: this corresponds to solving for retarded Green's functions, in accord with the usual prescription for calculating transport coefficients.  This simply involves imposing that only the negative sign solution in \eno{AlphaExponent} contributes.

The second boundary condition must be imposed not at the horizon, but at the boundary.  Each mode has two solutions near the boundary, one falling off more quickly and corresponding to a VEV deformation of the dual field theory, and the other falling off more slowly and corresponding to turning on a source.  Our prescription will be
to normalize the coefficient of the ``source" mode to unity.

Assuming we have imposed infalling boundary conditions, a part of the conserved flux \eno{AbelFlux} simplifies,
\eqn{ImPartHorizon}{
h \, {\rm Im} (X^* X') = - \omega |x_0|^2 e^{B(r_H) - A(r_H)} \,,
}
where we used the fact that $\alpha$ is pure imaginary. One can evaluate the conserved flux at any $r$, but since one boundary condition is at the horizon and the other at the boundary, one must in general solve for the mode everywhere in order to do so.  As we shall see, in some cases it is possible to do this analytically in the $\omega \to 0$ limit; in general, we shall solve the fluctuation equations numerically.

\subsection{Traceless graviton and shear viscosity}

This case is well-known, and the outcome, $\bar\eta/s = 1/4\pi$, is guaranteed by the general arguments of \cite{Buchel:2003tz,Kovtun:2004de}.  However, it is useful to go through an explicit calculation as a warmup for the more difficult conductivity and bulk viscosity cases, to be treated next.

Letting $Z$ be any of the traceless $h_{ij}$, the corresponding fluctuation equation is the massless scalar equation,
\eqn{ZhydroEqn}{
Z'' + \left( 4 A' - B' + {h' \over h} \right) Z'  + { e^{2B-2A}\over h^2} \, \omega^2 Z = 0 \,.
}
The behavior of $Z$ near the boundary is
\eqn{}{
Z = Z_{(0)} + \ldots + Z_{(4)} e^{-4 r/L} + \ldots \,,
}
and so the proper boundary condition is simply to take $Z_{(0)}= 1$.  The associated conserved flux from Abel's identity is
\eqn{ZFlux}{
{\cal F}_Z = h e^{4A-B} \, {\rm Im} \, (Z^* Z') \,,
}
and the Kubo formula for the corresponding transport coefficient, the shear viscosity, is
\eqn{}{
\bar\eta = - {1 \over 2 \kappa^2} \lim_{\omega \to 0} {1 \over \omega} {\cal F}_Z
\,.
}
The simplicity of the equation \eno{ZhydroEqn} in the $\omega \to 0$ limit allows one to solve for the shear viscosity analytically; we take our discussion from \cite{Gubser:2008sz}.
 In the $\omega \to 0$ limit  the term with no derivatives vanishes, and we are left with
\eqn{}{
\partial_r (\log Z') = -\partial_r (4A -B + \log h) \,,
}
which has the solution
\eqn{}{
Z = a_0 + b_0 \int_r^\infty dr\,  {e^{-4A+B} \over h} \,.
}
The second term is technically not allowed at strict $\omega =0$ due to a logarithmic divergence; it may be kept for very small $\omega$, but for us it is enough to note that matching to the near-horizon expansion
\eqn{}{
Z(r) \approx z_0 (r - r_H)^{\alpha } = z_0  (1 +\alpha  \log (r - r_H) + \ldots) \,,
}
we have $z_0= a_0$; however as $r \to \infty$ we see $a_0 = Z_{(0)}$.  Thus the boundary condition $Z_{(0)} = 1$ imposes $z_0 = 1$ near the horizon. Using \eno{ImPartHorizon} and \eno{ZFlux}, one can then evaluate $\bar\eta$ at the horizon,
\eqn{}{
\bar\eta = {1 \over 2 \kappa^2} e^{3 A(r_H)} \,,
}
which using \eno{TemperatureEntropy} implies for the shear viscosity to entropy density ratio the familiar universal result,
\eqn{}{
{\bar\eta \over s} = {1 \over 4 \pi} \,.
}
The other transport coefficients we consider will not be universal in this way; from a practical standpoint, imposing the boundary condition at infinity will not impose a universal constraint on the near-horizon behavior.  A similar technique to the one reviewed here will, however, be useful in calculating the conductivity of the one-charge ${\cal N}=4$ black hole of the next section.

\subsection{Gauge field, conductivity and diffusion}

Because of $SO(3)$ symmetry, we can treat each of the fluctuations $a_i$ separately: they cannot mix at linear order.  For notational simplicity, we will drop the index $i$ altogether and use $a$ to denote one of the components of $a_i$: for example, $a=a_1$ to compute the conductivity in the $x^1$ direction.  The fluctuation equation is
\eqn{aEqn}{
a'' + \left( 2A' - B' + {h' \over h} + { \phi' f'(\phi) \over f} \right)a' + {e^{-2A} \over h}  \left( {e^{2B} \over h} \, \omega^2 - \, f(\phi) {\Phi'}^2 \right)a = 0\,,
}
and the associated conserved flux is
\eqn{aFlux}{
{\cal F}_a = h f(\phi) e^{2A-B} \, {\rm Im} \, (a^* a') \,.
}
Near the boundary the gauge field fluctuation behaves as
\eqn{aFar}{
a = a_{(0)} + a_{(2)} e^{-2r/L} + \ldots \,,
}
so the appropriate boundary condition is simply $a_{(0)} = 1$.
The Kubo formula for this mode determines the (zero-frequency) conductivity $\lambda$,
\eqn{lambdaEqn}{
\lambda = -  {L^2 \over 2 \kappa^2} \lim_{\omega \to 0} {1 \over \omega} {\cal F}_a \,.
}
This quantity is related to the frequency-dependent complex conductivity, defined as
\eqn{Conductivity}{
\sigma \equiv - {i \over \omega} {L \over \kappa^2 } {a_{(2)} \over a_{(0)}}\,.
}
by
\eqn{}{
\lambda = \lim_{\omega \to 0} \Re \sigma(\omega) \,
}
which follows from \eno{aFlux} and \eno{Conductivity} using the boundary condition.
The diffusion constant $D$ is related to the conductivity by
\eqn{lambdaDchi}{
\lambda = D \chi \,,
}
where the susceptibility is
\eqn{}{
\chi \equiv \left( {\partial \rho \over \partial \mu} \right)_T \,,
}
leading to the the Nernst-Einstein relation,
\begin{eqnarray}
D \chi = \lim_{\omega \to 0} \Re \sigma(\omega) \,.
\end{eqnarray}
Unlike the shear viscosity, the conductivity does not take a universal value; notice that the equation \eno{aEqn} still has a zero-derivative term at $\omega  =0$.  For the example of the next section, however, we will be able to solve the $\omega \to 0$ limit and find an analytic solution for the conductivity.

\subsection{Scalar fluctuation and bulk viscosity}

Finally the scalar equation is
\eqn{SEqn}{
{\cal S}''& + \left(4 A' - B' + {h' \over h}\right) {\cal S}' + 
\left( {e^{2B-2A} \over h^2}  \omega^2+ \Sigma(r) \right) {\cal S} = 0 \,,
}
where $\Sigma(r)$ is determined by the background:
\eqn{}{
\Sigma(r) &\equiv
 {e^{-2A} \over 18 f h^2 {A'}^2 } \Big(-18 h {A'}^2 {f'}^2 {\Phi'}^2 - e^{2A} f h^2 {\phi'}^4 +6 f h A' \phi' \Big[ -2 e^{2A+2B} V'
 \cr &\qquad{}  + e^{2A} h' \phi' + f' {\Phi'}^2 \Big] 
 + 3 f {A'}^2 \left[  8 e^{2A} h^2 {\phi'}^2 + 3 h {\Phi'}^2 f'' - 6 e^{2A+2B} h V'' \right]
 \Big) \,.
}
The associated conserved flux is
\eqn{SFlux}{
{\cal F}_{\cal S} = h e^{4A-B} \, {\rm Im} \, ({\cal S}^* {\cal S}') \,,
}
and one may in principle calculate the bulk viscosity $\zeta$ from this flux using the formula
\eqn{}{
\zeta = - {2 \over 9 \kappa^2} \lim_{\omega \to 0} {1 \over \omega} {\cal F}_{\cal S} \,,
}
with ${\cal S}$ suitably normalized.
Since a gauge may be chosen where ${\cal S} = \tilde\phi$, it is evident that ${\cal S}$ has the same asymptotic behavior as the scalar $\phi$ itself,
\eqn{}{
 {\cal S}(r) = {\cal S}_{(4-\Delta)} e^{r(\Delta-4)/L} + \ldots + {\cal S}_{(\Delta)} e^{-r\Delta/L} + \ldots \,.
}
Thus to calculate \eno{SFlux}, one must properly normalize the coefficient ${\cal S}_{(4-\Delta)}$ of the dying exponential.  This is not as straightforward numerically as normalizing an asymptotic constant.

Since the bulk viscosity is a quantity extracted from the trace of the energy-momentum tensor, and the energy-momentum tensor couples to the graviton, it is the mode ${\cal H}$ that we more naturally wish to canonically normalize.
The scaling of ${\cal H}$ depends on $\phi(r)$ via \eno{HandS}.  If the background scalar profile includes the ``source" term $\phi_{(4-\Delta)}$, we have $A'/\phi' \sim e^{r(4 - \Delta)/L}$ and we end up with the ${\cal H}$ scaling
\eqn{HAsymptotic}{
{\cal H}(r) = {\cal H}_{(0)} + \ldots + {\cal H}_{(4 - 2 \Delta)} e^{r(4 - 2 \Delta)/L} + \ldots \,,
}
and the suitable boundary condition to calculate the bulk viscosity is ${\cal H}_{(0)} = 1$, an easier prescription to implement.\footnote{The asymptotic scaling of ${\cal H}$ is different when the background scalar profile only contains the VEV term $\phi_{(\Delta)}$ but no source.  However, in this case the conformal invariance of the dual field theory is not explicitly broken, and the bulk viscosity is identically zero.}

The fluctuation equation for ${\cal H}$ simplifies most when we use the background equations of motion to eliminate $V'$ and ${\phi'}^2$ in favor of $\phi''$ and $A''$:
\eqn{HEqn}{
{\cal H}'' + \left( 4A' -  B' + {h' \over h} + {2 \phi'' \over \phi'} - {2 A'' \over  A'} \right) {\cal H}' + 
 \left( {e^{2B-2A} \over h^2}  \omega^2+ \Sigma_{\cal H}(r) \right) {\cal H} = 0 \,,
 }
 with
  \eqn{}{
 \Sigma_{\cal H} = {h' \over h} \left({A'' \over A'} - {\phi''\over \phi'} \right) + {e^{-2A} \over h \phi'} \left( 3 A' f' - f \phi' \right) {\Phi'}^2 \,.
 }
 One can check that the boundary behavior as described above does indeed result from the asymptotic expansion of \eno{HEqn}, as it must.
The coefficient of the first-derivative term is again explicitly integrable, and it is then easy to use Abel's identity to show that the conserved flux is
\eqn{HFlux}{
{\cal F}_{\cal H} = {e^{4A-B} h {\phi'}^2 \over 4 {A'}^2} {\rm Im}\, ({\cal H}^* {\cal H}') \,,
}
where we chose the normalization so that this conserved flux coincides exactly with ${\cal F}_{\cal S}$ \eno{SFlux} when one  substitutes \eno{HandS}:
\eqn{}{
{\cal F}_{\cal H} = {\cal F}_{\cal S}\Big({\cal S} \to - {\phi' {\cal H} \over 2 A' }\Big)  \,.
}
Thus the bulk viscosity may most easily be calculated as
\eqn{BlkKub}{
\zeta = - {2 \over 9 \kappa^2} \lim_{\omega \to 0} {1 \over \omega} {\cal F}_{\cal H} \,,
}
with the boundary condition ${\cal H}_{(0)} = 1$.

\subsection{Large-$N_c$ counting}
\label{LargeNSec}

The various Kubo formulae are all proportional to the five-dimensional gravitational constant $1/\kappa^2$, since they all come ultimately from evaluating the five-dimensional action.  The gravitational constant can be expressed in terms of the AdS radius $L$ and the underlying string coupling $g_s$ and Regge slope $\alpha'$ when an explicit string theory construction is specified.  For $AdS_5 \times S^5$, the result is
\eqn{}{
{1 \over \kappa^2} = {L^5 \over 64 \pi^4 g_s^2 {\alpha'}^2} \,,
}
while the AdS radius is
\eqn{}{
L^4 = 4 \pi g_s N_c {\alpha'}^2\,,
}
with $N_c$ the number of colors in the dual gauge theory (the five-form flux from the gravity point of view).
Field theory quantities should not involve $\alpha'$, which drops out of the combination
\eqn{LKappa}{
{L^3 \over \kappa^2} = {N_c^2 \over 4 \pi^2} \,,
}
and all the Kubo formulae indeed result in this combination on dimensional grounds.  As a result, all three transport coefficients go like $N_c^2$ in the large-$N_c$ limit defined by the dual gauge theory.

For AdS/CFT models not based on a known string theory construction, the precise coefficient in \eno{LKappa} is not determined.  However, the $N_c^2$ dependence is expected to remain the same for any gravity dual of a four-dimensional gauge theory.  For the QCD-like holographic critical models of section~\ref{QCDBHSec}, we will use the formula \eno{LKappa} with the understanding that the overall normalization is not truly determined, but the $N_c^2$ factor is expected to be robust.  Note too that the factor \eno{LKappa} cancels out of $\zeta/s$ and $D$, and so will only be relevant for us in $\lambda/T$.

\section{One-charge ${\cal N}=4$ black hole}
\label{OneChargeSec}

Before discussing our primary interest, the numerical QCD-like  critical black holes, we first consider a family of black hole backgrounds where the thermodynamics are known analytically, which we call the one-charge ${\cal N}=4$ black hole \cite{Behrndt:1998jd, Gubser:1998jb, Kraus:1998hv, Cai:1998ji, Cvetic:1999ne, Cvetic:1999rb}.  These 
geometries are known solutions of string theory, coming most simply from a truncation of the maximally supersymmetric gauged supergravity in five dimensions, which is in turn a truncation of the dimensional reduction of type IIB string theory on $AdS_5 \times S^5$.  The dual field theory configurations should be thought of as states in ${\cal N}=4$ super-Yang-Mills theory with a temperature and a chemical potential for a $U(1)$ subgroup of the $SO(6)$ R-symmetry.  In the appendix, we discuss a second family, the two-charge ${\cal N}=4$ black hole, which does not display finite-temperature critical phenomena.

Since the temperature and chemical potential are the only massive parameters, dimensionless quantities can only depend on their ratio; for this reason the phase diagram is not truly two-dimensional, but is more properly thought of as depending on this single ratio.  The phase diagram has the form of a semi-infinite line, ending on a critical point.

Because the conformal invariance of the theory is not broken explicitly, the bulk viscosity is identically zero.  The conductivity and associated diffusion have been calculated by \cite{Maeda:2008hn, Son:2006em}.  We reproduce the calculation for two reasons.  First, since it is analytically solvable, it allows us to check our numerical methods against a known analytic solution; and second, it exhibits several features that will persist to the QCD-like black holes.

In this section, in keeping with the literature, we use $\mu$ to denote a parameter in the solutions, and thus use $\Omega$ for the chemical potential.  Also, it is inconvenient to set $B(r) =0$ for these solutions, so for this section (and the appendix) only we employ a radial coordinate where AdS space is related to that given in \eno{AdS} as $\exp(r_{\rm there}/L) = r_{\rm here}/L$.

\subsection{Thermodynamics and phase diagram of the one-charge ${\cal N}=4$ black hole}

The one-charge ${\cal N}=4$ black hole is a solution to the Lagrangian \eno{LwithF} with the potential and gauge kinetic function,
\eqn{}{
V(\phi) = - {1 \over L^2} \left( 8 e^{\phi \over \sqrt{6}} + 4 e^{- \sqrt{2 \over 3} \phi} \right) \,, \quad
f(\phi) = e^{-2 \sqrt{2 \over 3}\phi} \,.
}
 The solution
takes the form
\eqn{}{
A(r) &= \log {r \over L} + {1 \over 6} \log \left( 1 + {Q^2\over r^2} \right)\,, \quad
B(r) = - \log {r \over L}- {1 \over 3} \log \left( 1 + {Q^2\over r^2} \right)\,, \cr
h(r) &= 1- {\mu L^2 \over r^2(r^2 + Q^2) }\,, \quad
\phi(r) = - \sqrt{2\over 3} \log \left( 1 + {Q^2\over r^2} \right) \,, \quad
\Phi(r) = {\sqrt{\mu} Q \over r^2 + Q^2} -  {\sqrt{\mu} Q \over r_H^2 + Q^2}  \,,
}
and is  characterized by the charge parameter $Q$, mass parameter $\mu$ and the asymptotic AdS scale $L$.
The horizon is at 
\eqn{}{
r_H = \sqrt{{1 \over 2} \left(\sqrt{Q^4 + 4 \mu L^2} - Q^2 \right)} \,,
}
and in what follows we will generally trade the parameter $\mu$ for $r_H$.
The temperature and chemical potential can be expressed as
\eqn{}{
T = {Q^2 + 2 r_H^2 \over 2 \pi L^2 \sqrt{Q^2 + r_H^2}} \,, \quad \quad
\Omega = {Q r_H \over L^2 \sqrt{Q^2 + r_H^2}} \,,
}
For fixed $Q$ and $L$, $\mu$ is only required to be nonnegative; the limit $\mu \to 0$ corresponds to $r_H \to 0$ but the temperature approaches
\eqn{}{
T (\mu \to 0) \to{ Q \over 2 \pi L^2} \,. 
}
This value is properly thought of as the {\em limiting} temperature of a sequence of honest black holes with nonzero values of $\mu$.  The solution with strictly $\mu = 0$ is not  a black hole, as the horizon function is trivial $h(r) = 1$ and no horizon exists.  The $\mu = 0$ solution is supersymmetric and has been dubbed the ``superstar" by Myers and Tafjord \cite{Myers:2001aq}.

Choosing $Q$ to be positive along with $r_H$, $T$ and $\Omega$ are in general multivalued as functions of these parameters.  The transformation
\eqn{OneChargeTransform}{
r_H \to {Q\over 2} {\sqrt{Q^2 + 4 r_H^2} \over \sqrt{Q^2 + r_H^2}} \,, \quad \quad
Q \to r_H  {\sqrt{Q^2 + 4 r_H^2} \over \sqrt{Q^2 + r_H^2}} \,,
}
maps pairs $(r_H, Q)$ to physically distinct pairs with the same $(T, \Omega)$.  The transformation takes
\eqn{}{
{Q^2 \over 2 r_H^2} \to {2 r_H^2 \over Q^2} \,,
}
which defines a fixed locus,
\eqn{}{
 Q = \sqrt{2} r_H  \quad \to \quad  \pi T=  \sqrt{2} \Omega \,. 
}
Since dimensionless quantities like $s/T^3$ and $\chi/T^2$ will be functions only of the dimensionless ratio $\Omega/T$,  we should properly think of this system as a one-dimensional phase diagram with this locus interpreted as a fixed point $\Omega/T = \pi/\sqrt{2}$, rather than a fixed line.  The models of the next section add another massive parameter, with explicitly breaks conformal invariance and leads to a true two-dimensional phase diagram.

We notice that in general,
\eqn{}{
\pi^2 T^2 - 2 \Omega^2 = {(Q^2 - 2 r_H^2)^2 \over 4 L^4 (Q^2 + r_H^2)} \geq 0 \,,
}
which establishes the minimum value of $T$ for a given $\Omega$ lies at the fixed point,
\eqn{TBound}{
T \geq {\sqrt{2} \Omega \over \pi} \,.
}
Thus for $\pi T < \sqrt{2} \Omega$ there are no corresponding black holes, at the fixed point $\pi T = \sqrt{2} \Omega$ there is one, and for $\pi T > \sqrt{2} \Omega$ there are two.  The limit $\Omega \to 0$ produces two configurations dual under the transformation \eno{OneChargeTransform}: the superstar with $r_H=0$, $Q \neq 0$, and an uncharged black hole with $Q = 0$, $r_H \neq 0$.

Black holes on the two branches with the same $(T, \Omega)$ will not have the same values of other thermodynamic quantities, and so are physically distinct.  The entropy and charge density in terms of $Q$ and $r_H$ are
\eqn{}{
s = {2 \pi r_H^2 \over \kappa^2 L^3} \sqrt{Q^2 + r_H^2} \,, \quad \quad
\rho = {Q r_H \over \kappa^2 L^3}  \sqrt{Q^2 + r_H^2} \,,
}
which in terms of $T$ and $\Omega$ can be written 
\eqn{}{
s ={N_c^2  T^3 \over 16 \pi} (3 \pi \mp y)^2 (\pi \pm y)\,, \quad \quad
\rho ={N_c^2  T^2 \Omega \over 16 \pi^2} (3 \pi \mp y)^2 \,,
}
where the top sign corresponds to the branch with $\sqrt{2} r_H > Q$, the bottom sign  to the branch with $\sqrt{2} r_H < Q$, and where we have defined
\eqn{DefRoot}{
y\equiv \sqrt{\pi^2 - 2 \tilde\Omega^2} \,,
}
with $\tilde\Omega \equiv \Omega/T$.
Note that the positivity of the quantity inside the square root follows from \eno{TBound}.

The quantities $s$ and $\rho$ are the $T$ and $\Omega$ derivatives of the pressure, 
\eqn{}{
p =  {N_c^2 T^4\over16 \pi^2} \left(\pi^4+ 5 \pi^2  \tilde\Omega^2 - {\tilde\Omega^4 \over 2} \pm \pi y^3
\right) \,,
}
We note that the free energy density is $f = -p$, so for fixed $(T, \Omega)$ the $\sqrt{2} r_H > Q$ branch has a lower free energy and hence is thermodynamically preferred in the canonical ensemble.

The $U(1)$ susceptibility for the two branches is
\eqn{}{
\chi = {N_c^2 T^2 \over 8 \pi^2} \left( 5 \pi^2 - 3 \tilde\Omega^2 \mp 3 \pi y \pm {6 \pi \tilde\Omega^2 \over y} \right) \,,
}
which diverges along the fixed line for $y=0$.
The heat capacity and off-diagonal susceptibility diverge as well.  Thus the fixed point is a second-order phase transition.  
The determinant of the matrix of susceptibilities ${\cal S}$ is
\eqn{}{
\det {\cal S} =   {3N_c^4   T^4  \over 64 \pi} \left( 4 \pi^4 - 22 \pi^2  \tilde\Omega^2 + \tilde\Omega^4 \pm 4 \pi^3 y \pm { 46 \pi^3 \tilde\Omega^2 - 11 \pi \tilde\Omega^4 \over y} \right)  \,,
}
and while this is always positive for the thermodynamically preferred branch $\sqrt{2} r_H > Q$, it is always negative for the other; thus the non-preferred branch is thermodynamically unstable as well.

The exponents for this critical point have been calculated.  The fundamental relation is the approach of the density to the critical density, which for fixed $T$ takes the form:
\eqn{RhoCritical}{
\rho- \rho_c \sim |\Omega - \Omega_c|^{1/2} \equiv |\Omega - \Omega_c|^{1/\delta}\,,
}
defining the critical exponent
\eqn{}{
\delta = 2 \,.
}
An analogous relation holds for the approach to $T_c$ with fixed $\Omega$, as well as for 
the entropy.
The divergence of the conductivity $\chi$ (as well as the heat capacity $C_\Omega$) follows from (\ref{RhoCritical}),
\eqn{}{
\chi \sim |\Omega - \Omega_c|^{-1/2} \equiv |\Omega - \Omega_c|^{- \epsilon} \,,
}
where the exponent\footnote{In the literature for critical phenomena, the well-known exponent labels $\alpha$, $\beta$ and $\gamma$ are often associated to behavior approaching the critical point along the axis defined by the first-order line.  Since there is no first-order transition in this model, we avoid these exponent names. $\epsilon$ encodes the approach of $\chi$ to the critical point off the first-order axis.} $\epsilon \equiv 1- 1/\delta  = 1/2$.

\subsection{Conductivity and diffusion of the one-charge ${\cal N}=4$ black hole}

We may now turn to solving for the conductivity.  
First we consider the gauge field fluctuation equation \eno{aEqn}. This can be solved exactly at $\omega = 0$, allowing an analytic determination of the conductivity and the associated diffusion constant using similar techniques to those used  for the shear viscosity.  We find the $\omega = 0$ solutions
\eqn{aZeroSoln}{
a(r) &= C_1 {Q^2 + 2 r^2 \over (Q^2 + 2)(Q^2+ r^2)} + C_2 \, a_2(r) \,, 
}
where $a_2(r)$ is a more complicated expression including $\log(r - r_H)$ whose explicit expression is unenlightening.  In order to determine $C_1$ and $C_2$, one may match the zero-frequency solution \eno{aZeroSoln} to the near-horizon solution
\eqn{}{
a(r) = (r - r_H)^{\alpha} \,,
}
where the exponent $\alpha$ from \eno{AlphaExponent} corresponding to infalling boundary conditions is
\eqn{}{
\alpha = - {i \omega L^2 \sqrt{Q^2 + r_H^2 }\over 2 Q^2 + 4 r_H^2} \,.
}
Then looking at the near-boundary behavior, we can determine the series \eno{aFar} and solve for the frequency-dependent conductivity \eno{Conductivity}.  For small $\omega$ we find the result
\eqn{}{
\sigma = {i Q^2 \over 2 \omega \kappa^2 L } + {L (Q^2 + 2 r_H^2)^2 \over 8 r_H^2 \kappa^2 \sqrt{Q^2 + r_H^2}} + {\cal O}(\omega) \,,
}
which in terms of $T$ and $\Omega$ can be written
\eqn{}{
{\sigma  \over T} = {N_c^2 \over 4 \pi^2}  \left[ {i   \over 2 \tilde\omega} \left(2 \pi^2 - \tilde\Omega^2 \mp 2 \pi y\right)
+ {\pi^2 \over \pi \pm y}+ {\cal O}(\omega) \right]\,.
}
where $\tilde\Omega = \Omega/T$ and $\tilde\omega = \omega/T$.  The $1/\omega$ pole in ${\rm Im}\, \sigma$ indicates by the Kramers-Kronig relations a delta-function contribution to the real part, ${\rm Re}\, \sigma \propto \delta(\omega)$, not visible in our calculation.  This infinite contribution is characteristic of a translationally-invariant charged system \cite{Hartnoll:2008kx}.  Setting aside the delta-function, we have the zero-frequency conductivity \cite{Maeda:2008hn}
\eqn{}{
\lambda={N_c^2  T \over 4(\pi \pm y) } \,.
}
We also calculated this conductivity using the numerical methods that will be employed in the next section, and find excellent agreement.

Since $y \to 0$ at the critical point, the conductivity $\lambda$ approaches the constant
\eqn{LambdaCOneCharge}{ 
\lambda_c \equiv{N_c^2  T_c \over 4 \pi}    \,.
}
The conductivity not diverging is a sign of the large-$N_c$ domination of diffusive conduction (characteristic of model B) over  convective conduction (model H) \cite{Maeda:2008hn,Natsuume:2010bs}. However, the slope of $\lambda$ diverges as the critical point is approached.  One has
\eqn{LambdaExpOneCharge}{
\lambda - \lambda_c \sim (T - T_c)^{1/2} \,.
}
We note that the exponent is $1/\delta$; this transport coefficient has the same critical behavior as the thermodynamic quantities $s$ and $\rho$.
We will find an analogous  phenomenon for the QCD-like black holes.

The diffusion $D \equiv \lambda/\chi$ is given by
\eqn{}{
D= {4 \pi^2 \over T} {1 \over (\pi \pm y) (3 \pi \mp y) (3 \pi \mp y  \pm 4 \tilde\Omega^2/y)} \,. 
}
Since the conductivity is constant at the critical point, while the susceptibility diverges, the diffusion goes to zero.  On the stable branch at large temperature, $D$ asymptotes to its AdS value $D \to 1/2 \pi T$.  Because $\chi$ on the unstable branch is negative close to the critical point and passes through zero, the diffusion on the unstable branch is negative and diverges before returning from positive infinity.

\begin{figure}
  \centerline{\includegraphics[width=3.1in]{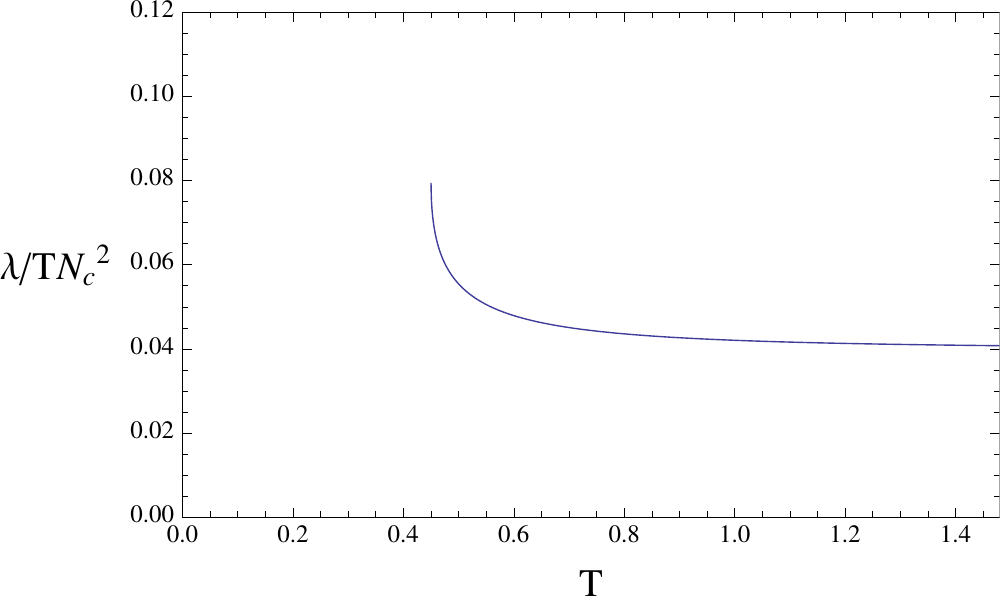}\quad\quad \includegraphics[width=2.9in]{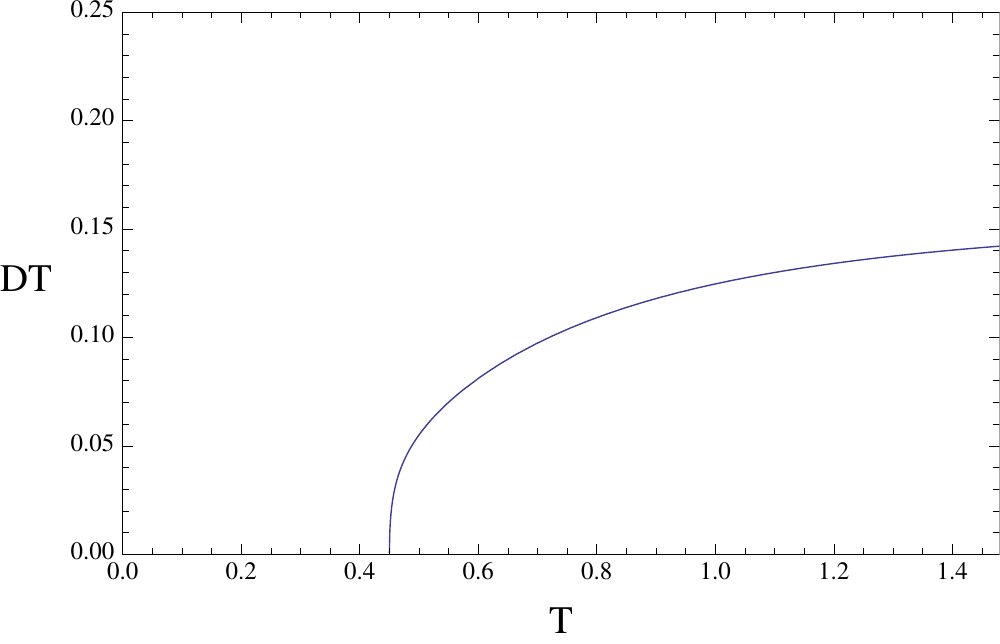}}
    \caption{The conductivity over temperature and diffusion times temperature for the one-charge black hole with $\Omega = 1$, for the stable branch only.  $\lambda/T$ approaches a finite value with an infinite slope at the critical point, while $DT$ goes to zero.}\label{OneChargeLambdaFig}
 \end{figure}

\begin{figure}
  \centerline{\includegraphics[width=3.1in]{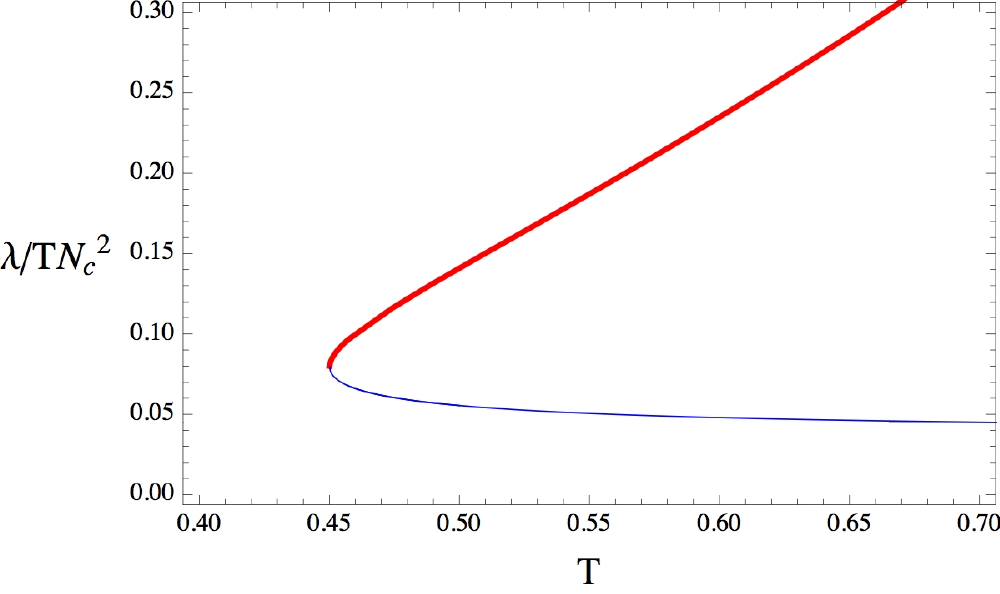}\quad\quad \includegraphics[width=2.9in]{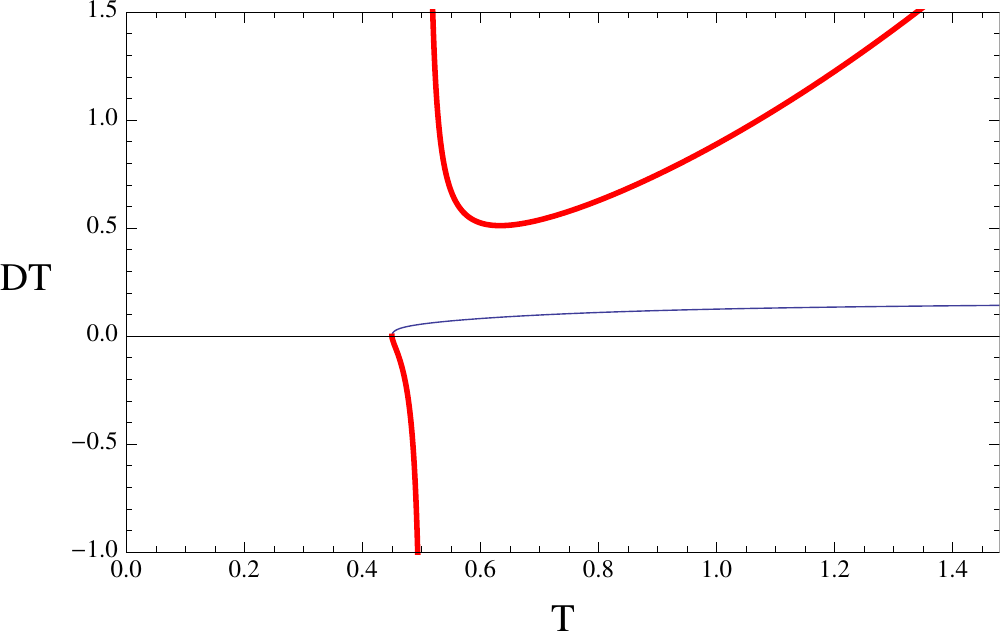}}
    \caption{The conductivity over temperature and diffusion times temperature for the one-charge black hole with $\Omega = 1$, for both branches; the unstable branch is the thick red line.  The approach to infinity at large $T$ is the superstar divergence.}\label{OneChargeLambdaAllFig}
 \end{figure}

Away from the critical point, an entirely distinct phenomenon occurs in the superstar limit.  Since as $y(\Omega \to 0) \to \pi$, the conductivity ratio $\lambda/T$ approaches zero at the charged black hole (the far end of the stable branch) and approaches infinity in the superstar limit (the far end of the unstable branch).  The rate of approach to infinity in the superstar limit becomes
\eqn{LambdaDoublePole}{
\lambda \sim {1 \over \Omega^2} \,.
}
We will find a similar divergence in the bulk viscosity of the holographic critical  black holes of the next section, away from the critical point and the region expected to be related to QCD.

\section{QCD-like holographic critical black holes}
\label{QCDBHSec}

We now turn to our primary interest, studying the transport coefficients of the class of solutions designed to emulate the thermodynamic and phase structure of QCD  \cite{DeWolfe:2010he}. These black holes are solutions to the Lagrangian ansatz \eno{LwithF} with the potential 
  \eqn{VChoice}{
  V(\phi) = {-12 \cosh \gamma\phi + b\phi^2 \over L^2} \qquad
    \hbox{with $\gamma = 0.606$ and $b = 2.057$} \,,
}
and the gauge kinetic function,
  \eqn{fChoice}{
  f(\phi) = {\sech\left[ {6 \over 5} (\phi-2) \right] \over \sech {12 \over 5}} \,.
}
The black hole solutions are of the form \eno{Background}.  For convenience the gauge $B(r) = 0$ was chosen, and the residual coordinate transformations \eno{Rescalings} and the freedom to shift $r$ were used to set
\eqn{CoordChoices}{
A(r_H) = 0 \,, \quad \Phi(r_H) = 0 \,, \quad h'(r_H) = 1/L \,, \quad r_H = 0\,.
}
The solutions were characterized by two initial conditions at the horizon,  $\phi_0 \equiv \phi(r_H)$ and $\Phi_1 \equiv \Phi'(r_H)$; the ensemble of black holes was generated by numerically integrating the equations \eno{SecondOrder} from the horizon to the boundary  for a large set of distinct $(\phi_0, \Phi_1)$, and the thermodynamic quantities of temperature $T$, entropy density $s$, chemical potential $\mu$ and baryon density $\rho$ were obtained for each.

\begin{figure}
\begin{center}
\includegraphics[scale=0.5]{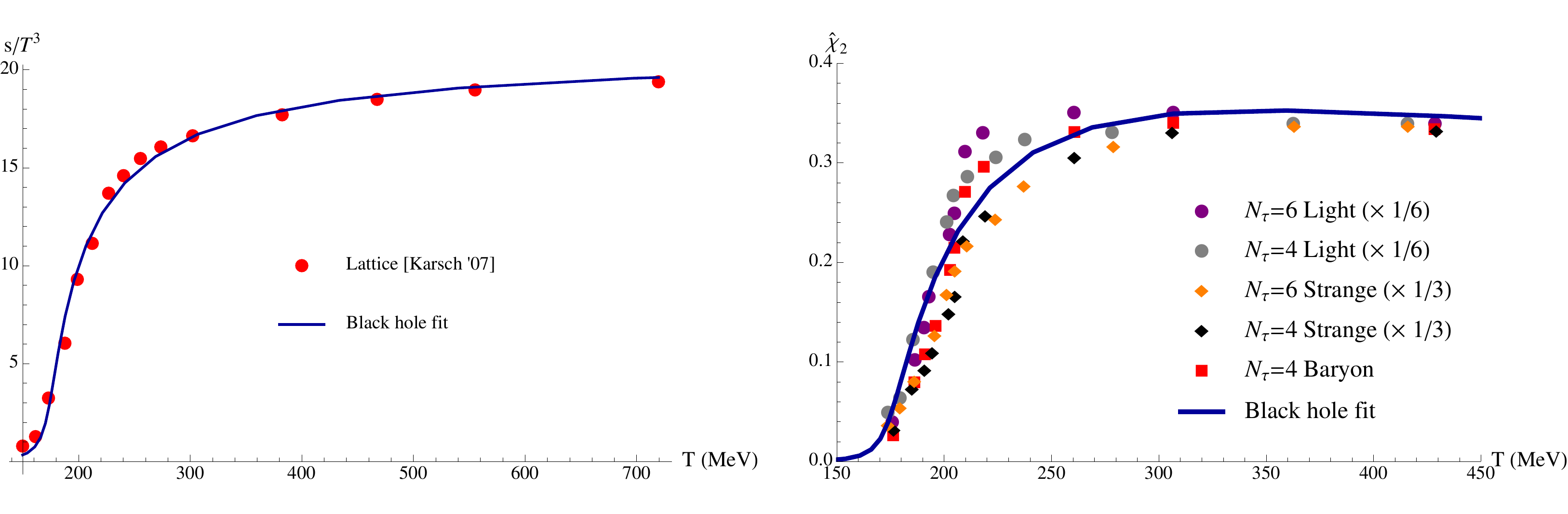}
\caption{The normalized entropy $s/T^3$ and quark susceptibility $\hat\chi_2 \equiv \chi_2/T^2$ at $\mu=0$, computed on the lattice and fit by black holes in the gravity theory defined by our choices of $V(\phi)$  and $f(\phi)$ (equations \eno{VChoice} and \eno{fChoice}).  Lattice data is taken from \cite{Karsch:2007dp}.
\label{ZeroMuComparison}}
\end{center}
\end{figure}

The functions \eno{VChoice} and \eno{fChoice} were chosen so that the thermodynamics of the $\mu=0$ black holes reproduce lattice results for the equation of state \cite{Gubser:2008ny} and the quark susceptibility \cite{DeWolfe:2010he}, building into the model the rapid crossover that connects the hadron phase to the quark-gluon phase in physical QCD with massive quarks; see figure~\ref{ZeroMuComparison}.  Normalizations for the scales of $T$ and $\mu$ were also derived from matching to the lattice results, allowing the thermodynamic quantities to be expressed in MeV.
The potential \eno{VChoice} leads to the scalar mass,
\eqn{}{
m^2_\phi L^2 \approx -0.293 \,,
}
with implies a dual operator that is barely on the relevant side of marginal,
\eqn{}{
\Delta_\phi \approx 3.93 \,,
}
designed to emulate the slow running of the QCD coupling constant somewhat above the confinement scale.  The profile of the scalar in the background includes a ``source" term that explicitly breaks conformal invariance.

At nonzero $\mu$ for this ensemble of black holes, the crossover sharpens into a line of first-order phase transitions ending at a critical point, as is expected for QCD.  The region of the first-order line is characterized by the existence of two distinct phases, realized as distinct black hole solutions at the same point in the phase diagram, as well as a third black hole corresponding to the thermodynamically unstable state lying in between the stable phases in the Maxwell construction. 

 The first-order line ends on the critical point.  For the static critical exponents defined as
 \eqn{CritDef}{
C_\rho  \equiv T \left( \partial s \over \partial T \right)_\rho &\sim |T - T_c|^{-\alpha} \,, \quad \quad \quad {\rm along \; first \; order \; axis}  \cr
\Delta \rho &\sim (T_c - T)^\beta \,, \quad \quad \quad {\rm along \; first \; order \; line}  \cr
\chi \equiv \left( \partial \rho \over \partial \mu \right)_T &\sim |T - T_c|^{-\gamma} \,, \quad \quad \quad {\rm along \; first \; order \; axis}  \cr
\rho - \rho_c &\sim |\mu - \mu_c|^{1/\delta} \,, \quad \quad \quad {\rm for}\ T = T_c \,,
}
this critical point was found numerically to have
\eqn{CritExpSummary}{
\alpha = 0 \,, \quad \quad
\beta \approx 0.482 \,, \quad \quad
\gamma \approx 0.942 \,, \quad \quad 
\delta \approx  3.035 \,,
}
with accuracies consistent with the Ising mean field values $(\alpha, \beta, \gamma, \delta) = (0, 1/2, 1, 3)$.  Matching the temperature and chemical potential scales to that of the lattice data, the location of the critical point was found to be
 \eqn{CriticalPosition}{
  T_c = 143 \, {\rm MeV} \qquad \mu_c = 783 \, {\rm MeV} \,.
 }
While these results generate a large phase diagram, the applicability to QCD is expected to be limited to a band surrounding the crossover and extending out into the plane, for two reasons.  First of all, the potential  \eno{VChoice} and gauge kinetic function  \eno{fChoice} were only matched to lattice data over a finite range.  In principle, one can imagine varying the functions so as to keep matching the data in the range required, while changing the results elsewhere.  The potential $V(\phi)$ was only matched up to about $\phi_0 \approx 7.5$, corresponding to $T_{\rm min} \approx 135\,{\rm MeV}$ at $\mu=0$; the crossover is at $T_c \approx 175\,{\rm MeV}$ on the $T$-axis.  Moreover, AdS/CFT is only expected to provide a good description of QCD in the region where no quasiparticle description is available; both above and below the crossover such a description is possible, with hadrons at lower $T$ and quarks and gluons at higher $T$.  Thus we expect our geometries to be reasonable potential models for QCD only near the crossover; and we hypothesize that this applicability extends along with the crossover out into the $T$-$\mu$ plane to the critical point.

\subsection{Boundary conditions}
\label{HorizonBoundaryCoords}

To calculate the correct transport coefficients (or thermodynamics for that matter) for these black holes, a technical detail must be taken into account.  The numerical integration beginning at the horizon produces solutions that asymptote to AdS space, but in general in coordinates that do not match standard AdS coordinates.  In the gauge $B(r) =0$, one usually uses the AdS coordinates \eno{AdS}, with $A(r) \to r/L$ and $h(r) \to 1 + {\cal O}(e^{-4r/L})$; however with the coordinate choices \eno{CoordChoices} used for the numerical solution,
the horizon function in general approaches a non-unit value $h \to h_0^{\rm far}$, and the warp factor goes like $A(r) \to r /(\sqrt{h_0^{\rm far}} L) + A^{\rm far}_0$.
To calculate thermodynamic and hydrodynamic quantities using the usual formulae, it is necessary to pass to a new set of coordinates that takes the usual AdS form, with unit normalization for the horizon function and $A \sim r/L$; this can be done by making a different rescaling \eno{Rescalings}.   

Moreover, the scalar field generically approaches $\phi \to \phi_A e^{-\nu A(r)}$, with $\nu \equiv 4 - \Delta_\phi \approx 0.07$.  It is useful to rescale the leading perturbation to the scalar to a standard magnitude $\phi_A = 1$, which can be achieved by adding a constant to $r$.    Fixing the value of $\phi_A$ is useful because turning on this perturbation corresponds to deforming the ultraviolet theory with an almost-marginal relevant operator, playing the role of the running coupling, and hence changing the theory.  We want every point on the phase diagram to correspond to the same theory, not  different theories, and hence we make $\phi_A$ identical in all of them.  Put another way, there are three massive parameters in the theory, $T$, $\mu$ and $\Lambda$, where $\Lambda$ is the scale associated with the scalar.  Normalizing $\phi_A = 1$ corresponds to measuring $T$ and $\mu$ in units of $\Lambda$.  The explicit breaking of conformal invariance coming from $\Lambda$ permits a non-zero bulk viscosity to be present.

The desired coordinates are achieved by the transformation
\eqn{}{
  \tilde{t} = \phi_A^{1/\nu} \sqrt{h^{\rm far}_0} \, t  \,, \quad \quad
  \tilde{\vec{x}} = \phi_A^{1/\nu} \vec{x}  \,,\quad \quad
  {\tilde{r} \over L}   =  {r \over \sqrt{h_0^{\rm far}}L} + A^{\rm far}_0 - 
    {\log(\phi_A^{1/\nu})} \,,
}
which is a combination of a scale transformation associated to $\phi_A^{1/\nu}$ and a time dilation by $\sqrt{h_0^{\rm far}}$ as well as a shift of $r$,
implying the relation of the functions,
\eqn{}{
  \tilde{A}(\tilde{r}) =A(r) - {\log(\phi_A^{1/\nu})}   \quad\quad
  \tilde{h}(\tilde{r}) = {1 \over h_0^{\rm far} }\, h(r)  \quad\quad
  \tilde\Phi(\tilde{r}) ={1 \over \phi_A^{1/\nu}  \sqrt{h_0^{\rm far}}} \, \Phi(r) \,.
}
Thermodynamic and hydrodynamic quantities are naturally calculated in terms of the tilded coordinates because of their standard near-AdS form. The black hole solutions exist in the untilded variables, but as we now show it is possible to compute certain ratios directly in the untilded coordinates. One may also use the thermodynamic formulae \eno{TemperatureEntropy}, \eno{MuRho} in the untilded variables to calculate ``horizon" quantities $T_H$, $\mu_H$, $s_H$ and $\rho_H$.  The true entropy and temperature are related to their ``horizon" counterparts as
\eqn{HorizonTs}{
T = {1 \over \phi_A^{1/\nu} \sqrt{h_0^{\rm far}}}\, T_H \,, \quad \quad
s = {1 \over \phi_A^{3/\nu} }\, s_H \,,
}
where the horizon quantities are constants independent of the black hole considered,
\eqn{}{
T_H = {1 \over 4 \pi L} \,, \quad \quad s_H = {2 \pi \over \kappa^2} \,.
}
Transforming from one set of coordinates to the other,
and using $\omega = \phi_A^{1/\nu} \sqrt{h_0^{\rm far}} \tilde\omega$, we have\footnote{Although under this rescaling one has $\tilde{a} = a/ \phi_A^{1/\nu}$, we assume the gauge field fluctuation is asymptotically normalized to one in both coordinate systems before being plugged into the Kubo formulae.}
\eqn{}{\bar\eta = {1 \over \phi_A^{3/\nu} }\, \bar\eta_H \,, \quad \quad 
\lambda = {1 \over \phi_A^{1/\nu}} \, \lambda_H \,, \quad \quad
\zeta = {1 \over \phi_A^{3/\nu} }\, \zeta_H \,.
}
Thus, we can assemble the invariants
\eqn{Invariants}{
{\bar\eta_H \over s_H} = {\bar\eta \over s} \,, \quad \quad 
{\lambda_H \over s_H^{1/3} } = {\lambda  \over s^{1/3}} \,, \quad \quad
{\zeta_H \over s_H} = {\zeta \over s} \,,
}
and thus these ratios can be calculated directly in the untilded variables.

With the gauge choices $B(r) = 0$ and \eno{CoordChoices}, the value of the exponent $\alpha$ characterizing the near-boundary behavior of the fluctuation equations is simply
\eqn{}{
\alpha = \pm i L \, \omega \,.
}
As usual, one should choose the minus sign to impose infalling boundary conditions at the horizon.  The boundary conditions at the boundary are that $a$ and ${\cal H}$ approach unity.  As mentioned previously, if one employed ${\cal S}$ instead of ${\cal H}$, it would be the coefficient of the $e^{-\nu r}$ term that had to be properly normalized; this is numerically more difficult.  The results outlined in the following subsections were obtained using the ${\cal H}$ equation, but the ${\cal S}$ method was checked as well, and while the ${\cal S}$ calculation developed numerical pathologies in certain (low-T) regions of the phase diagram, the two methods agreed well in the regions where both appeared reliable.

\subsection{Bulk viscosity for the QCD-like holographic critical black holes}
The computation of the bulk viscosity for the QCD-like black hole
solutions is a straightforward application of the Kubo formula
\eno{BlkKub}. To construct the mode solutions to the ${\cal H}$ fluctuation equation \eno{HEqn}
we used a near-horizon series expansion to seed a numerical integration routine which propagates the mode solution out to close to the boundary.

Because the  flux \eno{HFlux} satisfies Abel's identity, it must be
independent of the radial location at which it is evaluated; ascertaining that this holds is a check that the fluctuations obtained indeed satisfy \eno{HEqn}. After verifying that this remains true in
practice, we then evaluated the flux at a radial location $r_T$ where the
numerical solutions are particularly robust. The black hole solutions are defined from $r \sim 10^{-6}$ near the horizon out to $r \sim 10$, and we used a value around $r_T\sim 10^{-4}$.
With the fluctuation in hand, we normalized it to one asymptotically and inserted it into \eno{BlkKub} and used \eno{HorizonTs} and \eno{Invariants} to calculate the bulk viscosity over entropy density ratio $\zeta/s$.\footnote{Since $\bar\eta/s$ is a constant, $\zeta/s$ is simply proportional to $\zeta/{\bar\eta}$.}

\begin{figure}
  \centerline{\includegraphics[width=5in]{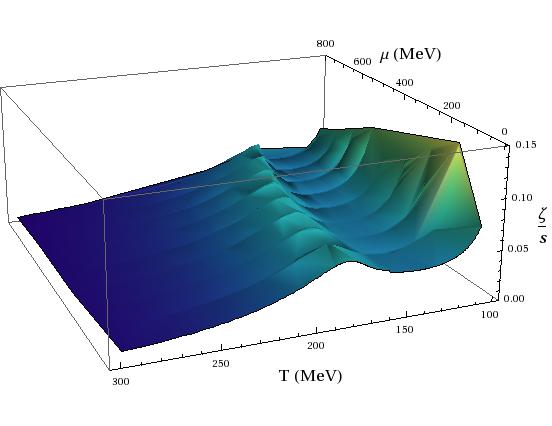}}
    \caption{The bulk viscosity over entropy density over the $T$-$\mu$ plane for the QCD-like black hole solutions.}\label{QCDZetaFig1}
 \end{figure}

\begin{figure}
  \centerline{ \includegraphics[width=3in]{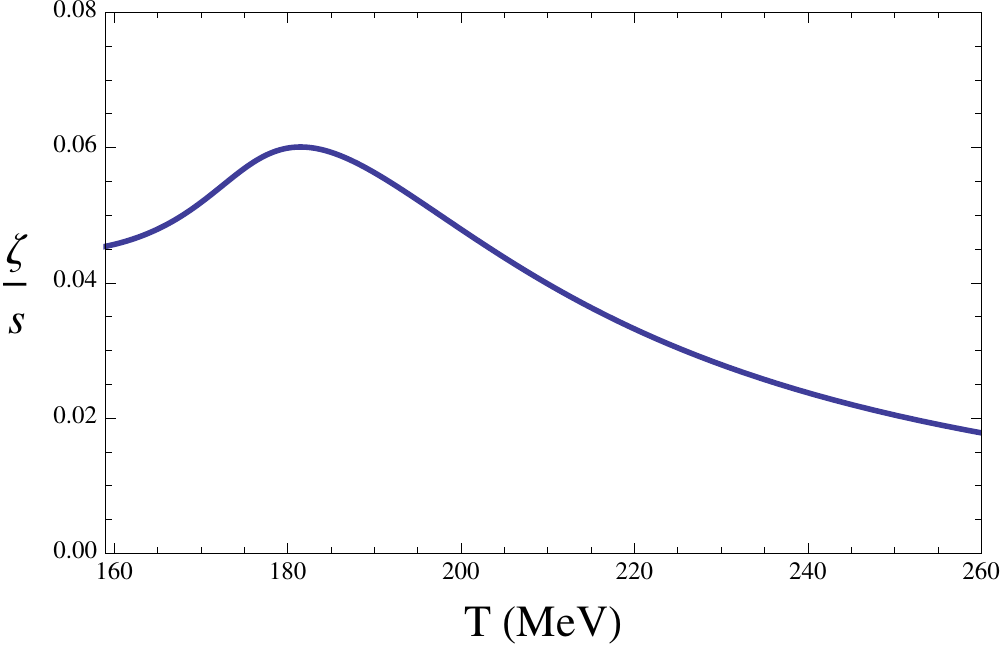} \quad \quad \includegraphics[width=3in]{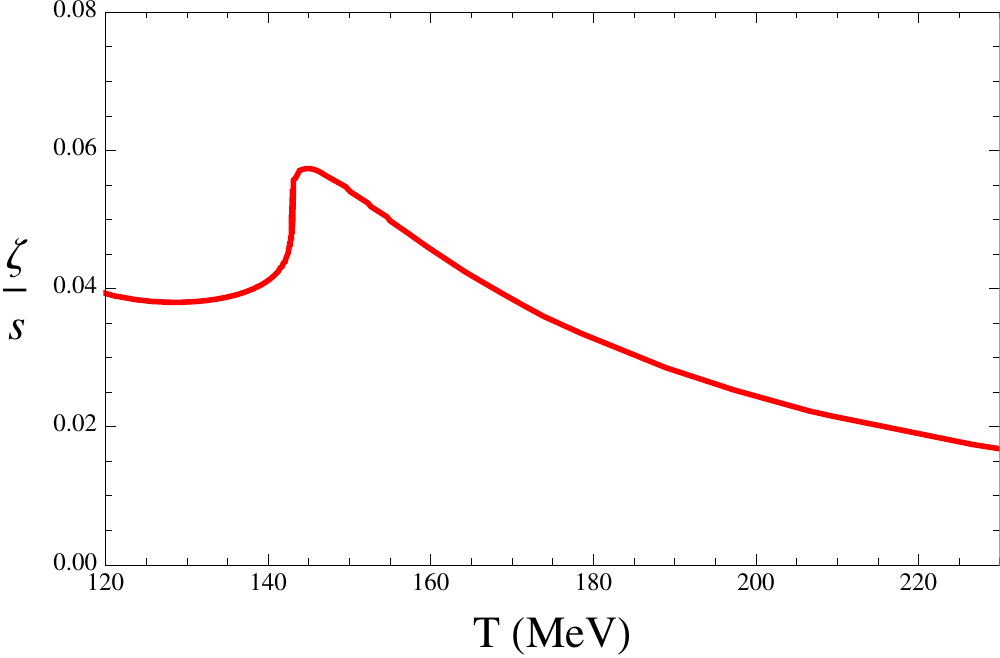}}
    \caption{The bulk viscosity over entropy density at zero chemical potential and at $\mu = \mu_c$ for the QCD-like black hole solutions; the ``bump" at left evolves into the peak with singular slope on one side as $\mu$ is increased.}\label{QCDZetaFig2}
 \end{figure}

The results of this calculation, when carried out for many points
on the $T$-$\mu$ plane, are shown in figure~\ref{QCDZetaFig1} up to $\mu = \mu_c$; in all figures in this section we have used the normalization of \cite{DeWolfe:2010he} to express $T$ and $\mu$ in MeV. Two features are immediately apparent.  One is the propagation of a ``bump" on the $T$-axis at the location of the crossover, out into the plane towards the critical point.  As the bump moves into the plane, it becomes increasingly asymmetrical as the slope on the lower-$T$ side becomes more vertical.  Precisely at the critical point, this slope diverges.  Constant-$\mu$ plots of $\zeta/s$ for both $\mu =0$ and $\mu = \mu_c$ are given in figure~\ref{QCDZetaFig2}.  The result at vanishing $\mu$ was obtained previously in \cite{Gubser:2008sz}.
 For $\mu > \mu_c$, the peak ``tips over" and the plot of $\zeta/s$ becomes multivalued.

 The first and more important conclusion from this critical behavior is that the bulk viscosity remains finite at the critical point; it does not diverge as is predicted in a number of models such as that of Onuki \cite{Onuki:1997}, given in equation \eno{Onuki}; see also \cite{Buchel:2009mf}.
 
Moreover, the way in which the quantity stays finite is interesting.  Approaching the critical point along the ``crest" of the peak turns out to be equivalent to approaching along the axis defined by the first-order line (but on the other side of the critical point); along this path, the value of $\zeta/s$ scarcely changes at all.  On the other hand, approaching the critical point from a direction other than the first-order axis, for example constant $\mu = \mu_c$ as depicted on the right of figure~\ref{QCDZetaFig2}, $\zeta/s$ develops a divergent slope; the same is true for constant $T$ or any other direction of approach, save only the first-order axis, where no such divergent slope is seen.

 The bulk viscosity is not the only quantity with this property.  It is common for thermodynamic functions to behave differently depending on whether the approach to the critical point is on the first-order axis or not.  In \cite{DeWolfe:2010he} it was found that the thermodynamic densities 
$s$ and $\rho$ also behave smoothly approaching the critical point along the first-order axis, but have divergent slope approaching off-axis.  Since the derivatives of these densities are the specific heat and susceptibilities, these properties are characterized by various static critical exponents: the smooth approach along the first-order axis corresponds to the vanishing $\alpha =0$, while the divergent slope approaching off the axis is encoded in the value of $\delta \approx 3$, giving an approach of the susceptibility {\em off} the first-order axis as
\eqn{ChiEpsilon}{
\chi \sim |\mu - \mu_c|^{-\epsilon} \sim |T - T_c|^{-\epsilon} \,, \quad \quad \quad {\rm off \; first \; order \; axis} \,,
}
with \eqn{}{
\epsilon \equiv 1 - {1 \over \delta} \approx {2/3} \,.
}
A natural guess is then that this behavior can be explained by the hypothesis that the bulk viscosity is a smooth function of the two densities: $\zeta= \zeta(s, \rho)$. Let us characterize the divergent slope of $\zeta$ by an exponent $\delta_\zeta$:
\eqn{}{
\zeta - \zeta_c  \sim |\mu - \mu_c|^{1/\delta_\zeta} \sim |T - T_c|^{1/\delta_\zeta} \,.
}
Then we would expect $\delta_\zeta = \delta$.  We analyzed the results of numerics and found an exponent consistent with this interpretation, although the strong sensitivity to the precise value of $\zeta_c$, which is difficult to determine due to the infinite slope, gave an uncertainty in the exponent in the range of $0.1 < 1/\delta_\zeta < 0.7$.  A similar argument can be carried through more precisely for the zero-frequency conductivity $\lambda$ of the one-charge ${\cal N}=4$ black hole.  In this case, the result \eno{LambdaExpOneCharge} shows that $\delta_\lambda = 2$, in agreement with $\delta=2$ for the one-charge model.

\begin{figure}
  \centerline{\includegraphics[width=3in]{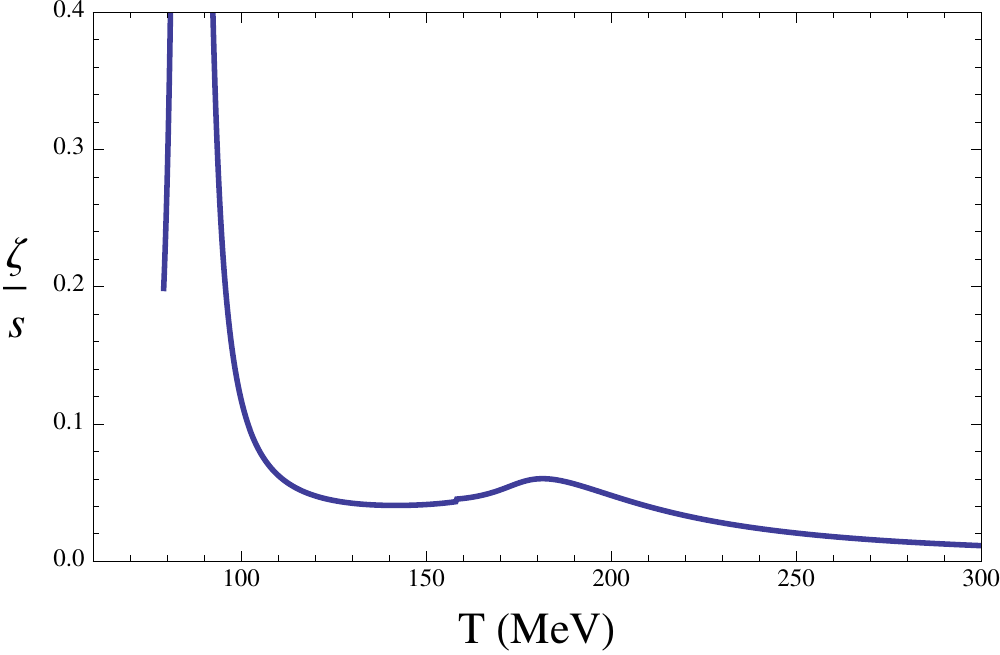}}
    \caption{The bulk viscosity over entropy density at zero chemical potential down to lower temperatures, showing the divergence.}\label{QCDZetaCrssovrFig}
 \end{figure}

The second salient feature of the $\zeta/s$ on the phase diagram is the 
strong rise at colder temperatures and smaller chemical potential. This behavior is shown for zero chemical potential
in figure \ref{QCDZetaCrssovrFig}, which shows the same thing as the left-hand-side of figure~\ref{QCDZetaFig2} but including lower values of $T$.  The bulk viscosity actually diverges at a temperature $T_*$, which on the T-axis is at $T_*(\mu = 0)\approx
86$~MeV, before becoming finite again at smaller temperatures; this divergence  extends from
the $T$-axis out along a curve into the finite chemical potential region of the
phase diagram.  The divergence is power law, with
\eqn{}{
\frac{\zeta}{s} \sim \left(T-T_*\right)^{-2} \,.
}
Notably, the thermodynamic densities $s$ and $\rho$ and their derivatives do nothing special at this locus.

From the gravity point of view, this divergence can be understood as related to a formation of a node in the mode solution for $\mathcal{H}(r)$.  For $T > T_*$, the fluctuation ${\cal H}(r)$ never crosses zero as it asymptotes to a constant value at the boundary.  For $T < T_*$ on the other hand, the solutions have a single zero before asymptoting to a constant.  As $T \to T_*$ from below, this node moves out towards the boundary, and precisely at $T = T_*$ the solution goes to zero at infinity; that is, the leading constant ${\cal H}_{(0)}$ in the asymptotic expansion of ${\cal H}$ \eno{HAsymptotic} vanishes.  Since the AdS/CFT prescription is to normalize this constant to unity, the scaling required to accomplish this normalization leads to the divergence.

A reasonable expectation is that for $T<T_*$, there is a perturbative instability in the black hole.  Because the thermodynamics is perfectly well-defined and stable in the vicinity of $T=T_*$, such an instability, if present, would be a decisive counter-example to the correlated stability conjecture of \cite{Gubser:2000ec,Gubser:2000mm}.

On the field theory side, however, it is unclear what this divergence means, and what the conjectured instability would be for $T<T_*$.  Importantly, this particular feature is not a prediction for QCD-like systems, since it lies well outside the range that was matched to lattice QCD data; $T_* \approx 86$ MeV is well below the lower bound of $T_{\rm min} \approx 135$ MeV where the matching to lattice QCD ended.  (In fact, the value of the temperature where $\zeta/s$ begins to rise towards the divergence is close to $T_{\rm min}$.)  Instead one should view this phenomenon as 
 an indication of what
features are generally possible in gauge/gravity transport. 

Interestingly, this divergence is reminiscent of the superstar divergence in the conductivity for the one-charge ${\cal N}=4$ black hole \eno{LambdaDoublePole}.  In both cases, the transport coefficient has a double pole in the thermodynamic variable as the divergence is approached.

\subsection{Conductivity and diffusion  for the QCD-like holographic critical black holes}

\begin{figure}
  \centerline{\includegraphics[width=5in]{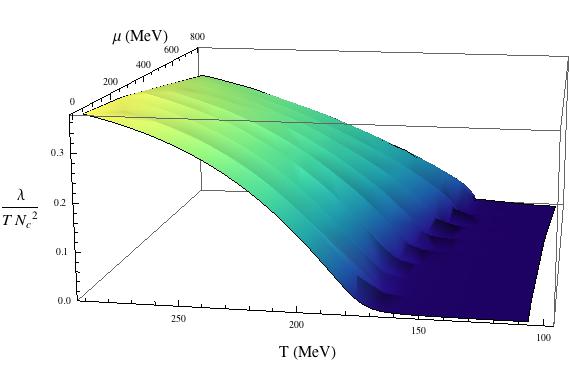}}
    \caption{The conductivity over temperature over the $T$-$\mu$ plane for the QCD-like black hole solutions.}\label{QCDlambdaFig1}
 \end{figure}

Utilizing a numerical algorithm identical in spirit to that employed
for the computation of the bulk viscosity, it was also possible to
construct solutions to the fluctuation equation for $a(r)$ \eno{aEqn} and
subsequently evaluate the conductivity Kubo formula
\eno{lambdaEqn}. The results of this calculation are displayed in figure~\ref{QCDlambdaFig1}, where the dimensionless ratio $\lambda/T$ is plotted over the $T$-$\mu$ plane.  Unlike the ratio $\zeta/s$, this quantity grows in the larger-$T$ region of the diagram.  Note that as mentioned in section~\ref{LargeNSec}, the overall normalization for this ratio is not known due to the lack of a string theory embedding, but the $N_c^2$ dependence should be general, so we plot $\lambda/T N_c^2$ using \eno{LKappa} as a concrete choice.

 \begin{figure}
\begin{center}
  \includegraphics[width=3.0in]{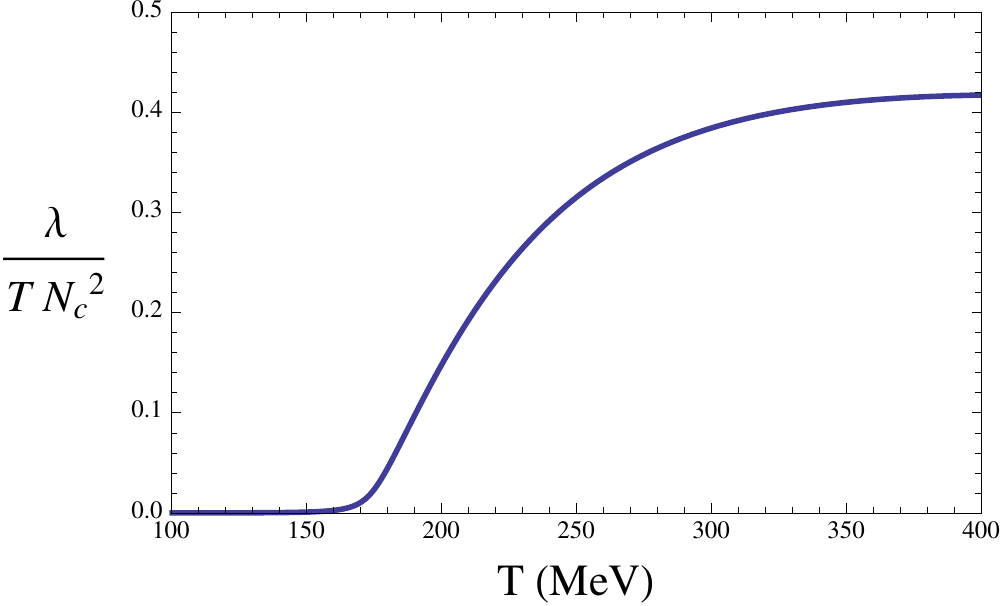}
 \quad \quad
  \includegraphics[width=3.0in]{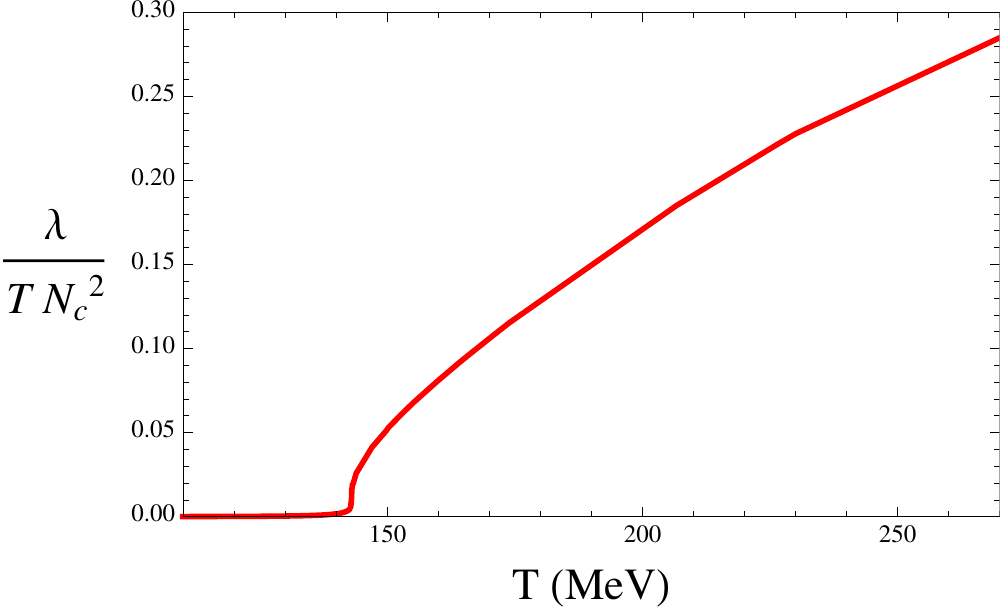} \end{center}
    \caption{Conductivity over temperature at zero chemical potential and at $\mu = \mu_c$ for the QCD-like black hole solutions; the smooth rise on the left evolves to the singular jump on the right as $\mu$ is increased. }\label{QCDlambdaFig2}
 \end{figure}

Once again, the most salient attribute is the propagation of the crossover feature out to the critical point, which is the sudden rise of $\lambda/T$.  Cuts at $\mu = 0$ and $\mu = \mu_c$ are displayed in figure~\ref{QCDlambdaFig2}.  As with the bulk viscosity, the conductivity attains a finite value at the critical point; and also like the bulk viscosity, the slope of the rise increases as the critical point is approached, until it diverges, except along the first-order axis.  Thus it is again natural to hypothesize that $\lambda = \lambda(s, \rho)$, and that off the first-order axis the conductivity approaches its critical value with exponent $1/\delta_\lambda$ 
\eqn{}{
\lambda - \lambda_c \sim |\mu - \mu_c|^{1/\delta_\lambda}\sim |T -T_c|^{1/\delta_\lambda} \,,
}
with $\delta_\lambda = \delta$.

As described in the introduction, the finiteness of the critical conductivity is reminiscent of dynamic critical model B rather than model H \cite{Hohenberg:1977ym}, matches the result for the one-charge ${\cal N}=4$ models (see equation \eno{LambdaCOneCharge} and \cite{Maeda:2008hn, Son:2006em}), and is consistent with the interpretation of a large-$N_c$ enhancement of diffusive conductivity over convective conductivity \cite{Natsuume:2010bs}.

We may also consider the baryon diffusion $D_B \equiv \lambda/\chi$. Since the conductivity diverges at the critical point, the finite critical conductivity implies
 vanishing baryon diffusion.  In \cite{DeWolfe:2010he}, the behavior of
the baryon susceptibility was  studied as the second order
point was approached along the first order line axis, and along directions
off this axis. From \eno{CritExpSummary} and \eno{ChiEpsilon}, one has that on and off the axis
the susceptibility diverges with
critical exponents $\gamma\approx 1$ and $\epsilon\approx 2/3$,
respectively. Thus, we find the rates of approach to zero for the diffusion
\eqn{}{
D_B \sim |T-T_c|^\gamma  \sim |T-T_c| \,, \quad\quad\quad \textrm{along first order axis} \,,
}
and
\eqn{}{
D_B \sim |T-T_c|^\epsilon \sim  |T-T_c|^{2/3}\,, \quad\quad\quad \textrm{off first order axis} \,.
}
\begin{figure}
\centering
  \includegraphics[width=3in]{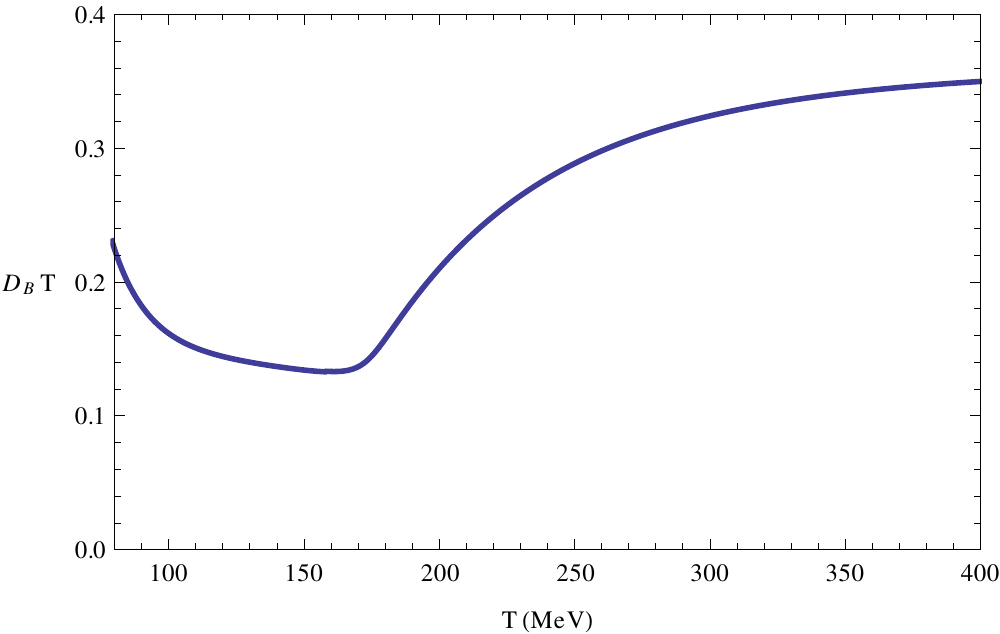}
    \caption{Baryon diffusion at zero chemical potential for the QCD-like black hole solutions.}\label{QCDDiffCrssovrFig}
 \end{figure}%
In figure \ref{QCDDiffCrssovrFig}, the conductivity divided by
temperature is plotted for vanishing chemical potential. In this
region of the phase diagram, the baryon susceptibility is particularly
easy to compute, as it can be explicitly related to the background metric and scalar \cite{DeWolfe:2010he}. As a
result, it is straightforward to extract the baryon diffusion along the
$T$-axis.
The diffusion also
experiences a rise at low temperature, but we are unable to
determine whether or not this rise culminates in a divergence (as is
the case for the bulk viscosity) as it manifests in a numerically
unreliable region of the phase diagram.

 \section{Conclusions}
 \label{ConclusionSec}

The primary lesson so far of the study holographic critical phenomena has been the suppression of fluctuations and convective transport by large $N_c$.  The models considered here have a thermodynamic crossover designed to emulate QCD, and correspondingly a critical point arises in the $T$-$\mu$ phase diagram.  However, rather than exact 3D Ising static critical exponents, we found mean field Ising critical exponents; and rather than divergences in the transport coefficients, we here find finite critical behavior. (In fact, one can view the famous constancy of the shear viscosity over entropy density, which holds throughout the phase diagram, as being the first example of this phenomenon, since mode-coupling theories predict a weak divergence.)  These results can be explained by the hypothesis that large-$N_c$ suppresses quantum corrections as well as  convective transport \cite{Natsuume:2010bs}. 

 Physical QCD of course has $N_c = 3$, but large-$N_c$ counting is still a useful way of looking at the theory for a number of phenomena.  Understanding better the extent to which the suppression phenomena that emerge from the AdS/CFT duals apply to real QCD is the primary outstanding question: in other words, could large-$N_c$-suppression in real QCD significantly push the behavior of physical quantities toward mean field and model B expectations, except very close to the critical point?  Understanding both the nonzero-momentum fluctuations and the finite-$N_c$ corrections to these models is essential for making further progress, and we hope to report on these issues in the future.

  \section*{Acknowledgments}

We are grateful to Alex Buchel, Tom DeGrand, Anna Hasenfratz, Andreas Karch, Subir Sachdev, and Amos Yarom for helpful discussions.
The work of O.D.\ and C.R.\ was supported by the Department of Energy under Grant
No.~DE-FG02-91-ER-40672.  The work of S.S.G.\ was supported by the Department of Energy under Grant No.~DE-FG02-91ER40671.

 \section*{Appendix: Two-charge ${\cal N}=4$ black hole}

Along with the one-charge ${\cal N}=4$ black hole solution of section~\ref{OneChargeSec}, another family of black holes coming from string theory is the so-called two-charge ${\cal N}=4$ solution, which has equal charges for two $U(1)$ gauge fields inside the $SO(6)$ R-symmetry.  Keeping only the diagonal gauge field, this geometry
solves our ansatz with potential and gauge kinetic function,
\eqn{}{
V(\phi) = - {1 \over L^2} \left( 8 e^{\phi \over \sqrt{6}} + 4 e^{- \sqrt{2 \over 3} \phi} \right) \,, \quad \quad
f(\phi) = e^{ \sqrt{2 \over 3}\phi} \,,
}
where the scalar potential matches that for the one-charge case, while the gauge kinetic function is slightly different.  The solution takes the form
\eqn{}{
A(r) &= \log {r \over L} + {1 \over 3} \log \left( 1 + {Q^2\over r^2} \right)\,, \quad
B(r) = - \log {r \over L}- {2 \over 3} \log \left( 1 + {Q^2\over r^2} \right)\,, \cr
h(r) &= 1- {\mu L^2 \over (r^2 + Q^2)^2 }\,, \quad
\phi(r) =  \sqrt{2\over 3} \log \left( 1 + {Q^2\over r^2} \right) \,, \quad
\Phi(r) = {\sqrt{2 \mu} Q \over r^2 + Q^2}  -  {\sqrt{2 \mu} Q \over r_H^2 + Q^2} \,.
}
This solution again is characterized by a charge $Q$, a mass parameter $\mu$ and the asymptotic AdS scale $L$.   The solution has a horizon at
\eqn{}{
r_H = \sqrt{L \sqrt{\mu} - Q^2} \,,
}
For fixed $Q$ and $L$, the mass parameter $\mu$ is bounded below by
\eqn{}{
\mu \geq \mu_{min} \equiv {Q^4 \over L^2} \,,
}
and at $\mu = \mu_{min}$ we have $r_H = T = 0$; extending $\mu < \mu_{min}$ results in a naked singularity.

It is more convenient to trade the parameter $\mu$ for $r_H$, which runs over all nonnegative values.
The temperature and chemical potential are
\eqn{}{
T = {r_H \over \pi L^2} \,, \quad \quad
\Omega = {\sqrt{2} Q \over L^2} \,,
}
giving a one-to-one relationship between the black hole parameters ($r_H$, $Q$) and the thermodynamic parameters ($T$, $\Omega$); a single black hole exists at each point in the phase diagram, and hence a single phase.  The entropy and charge density are
\eqn{}{
s &=  {2 \pi r_H \over \kappa^2 L^3} (Q^2 + r_H^2) = {N_c^2 T \over 4} 
    (2 \pi^2 T^2 + \Omega^2) \,, \cr 
\rho &= {\sqrt{2} \over \kappa^2 L^3} Q (Q^2 + r_H^2)  = {N_c^2 \Omega \over 8 \pi^2}
    (2 \pi^2 T^2 + \Omega^2) \,,
}
which can be obtained as derivatives of the pressure
\eqn{}{
p = {N_c^2 \over 32 \pi^2} (2 \pi^2 T^2 + \Omega^2)^2 \,.
}
The $U(1)$ susceptibility is
\eqn{TwoChargeChi}{
\chi \equiv \left( \partial \rho \over \partial \Omega\right)_T= {N_c^2  \over 8 \pi^2}(2 \pi^2 T^2 + 3 \Omega^2) \,,
}
and it along with other derivatives of $s$ and $\rho$ such as as the specific heat are everywhere well-behaved; this example has no phase transitions.

\begin{figure}
  \centerline{\includegraphics[width=3.1in]{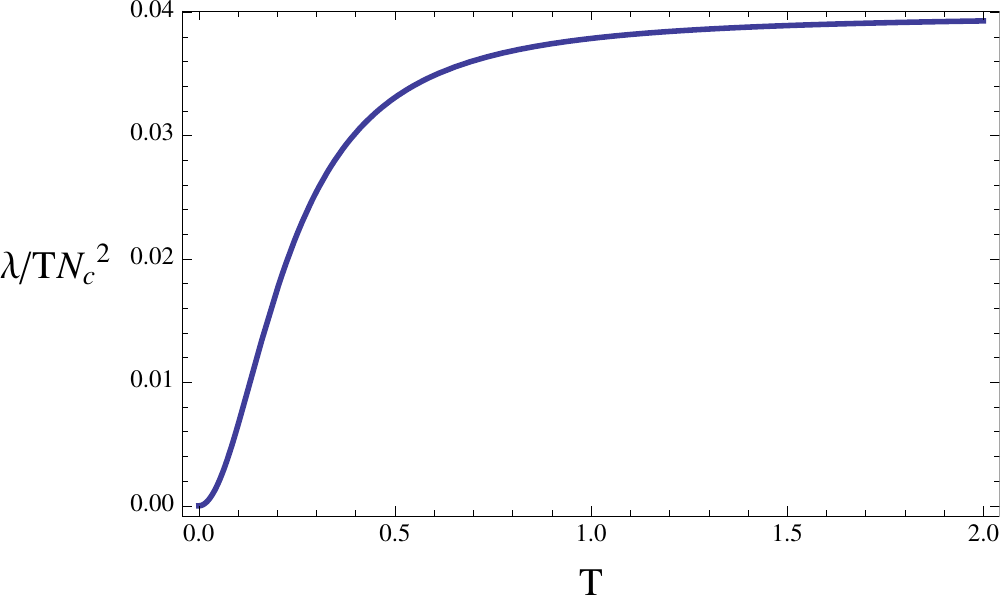}\quad\quad \includegraphics[width=2.9in]{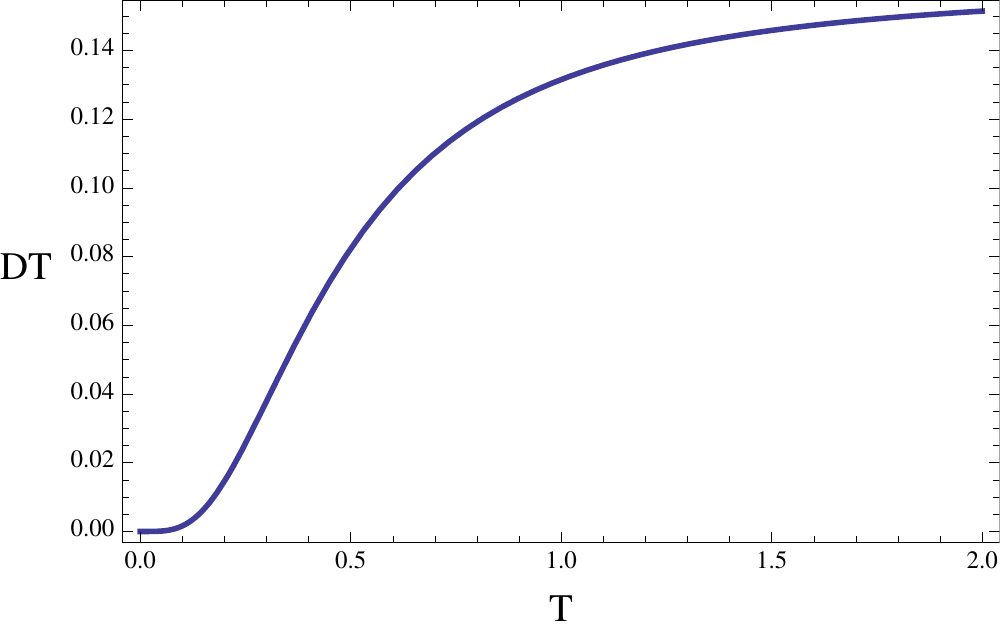}}
    \caption{The conductivity over temperature and diffusion times temperature for the two-charge black hole with $\Omega = 1$.}\label{TwoChargeFig}
 \end{figure}

We can consider  the conductivity and related diffusion coefficient for this model.
Proceeding analogously to the one-charge case, we find the $\omega = 0$ solutions
\eqn{}{
a(r) &= C_1 {r^2 \over Q^2+ r^2} + C_2 \, a_2(r) \,, 
}
where, as for the one-charge case, $a_2(r)$ is a more complicated expression including $\log(r - r_H)$.  Matching to the near-horizon solution
\eqn{}{
a(r) = (r - r_H)^{\alpha} \,,
}
with the exponent $\alpha$ imposing infalling boundary conditions given as \eno{AlphaExponent}
\eqn{TwoAlpha}{
\alpha = - {i \omega L^2 \over 4 r_H} \,,
}
we determine $C_1$ and $C_2$ and solve for the conductivity \eno{Conductivity}.  For small $\omega$ we find the result
\eqn{}{
\sigma(\omega) = {i Q^2 \over \omega \kappa^2 L} + {r_H^3 L \over 2 \kappa^2 (Q^2 + r_H^2)}  + {\cal O}(\omega) \,.
}
It is straightforward to convert this to dependence on the thermodynamic quantities $T$, $\Omega$, and one finds
\eqn{TwoChargeSigma}{
 {\sigma \over T} = {N_c^2 \over 4 \pi^2}  \left( {i {\tilde\Omega}^2 \over 2 \tilde\omega} + {\pi^3 \over 2 \pi^2 + \tilde\Omega^2} +  {\cal O}(\omega)\right) \,,
}
where again $\tilde\Omega = \Omega/T$, $\tilde\omega = \omega/T$.  As in the one-charge case we find a $1/\omega$ pole in the imaginary part, indicating a delta function in the real part, and
the remaining zero-frequency conductivity
\eqn{}{
\lambda ={N_c^2 \pi T^3 \over 4( 2 \pi^2 T^2 + \Omega^2) }  =
  {N_c^2  \pi T  \over 4( 2 \pi^2 + \tilde\Omega^2)} \,,
}
which using the susceptibility \eno{TwoChargeChi} gives the diffusion constant
\eqn{}{
D = {1 \over 2 \pi T } \left( 1 \over 1 + 2\tilde\Omega^2 /\pi^2+ 3\tilde\Omega^4/4 \pi^4 \right) \,.
}
We plot $\lambda/T N_c^2$ and $D T$ in figure~\ref{TwoChargeFig}.

\bibliographystyle{JHEP}
\bibliography{Fluctuations}

\end{document}